\documentclass[fleqn,usenatbib]{mnras}
\usepackage[T1]{fontenc}
\DeclareRobustCommand{\VAN}[3]{#2}
\let\VANthebibliography\thebibliography
\def\thebibliography{\DeclareRobustCommand{\VAN}[3]{##3}\VANthebibliography}

\usepackage{graphicx}	
\usepackage{amsmath}	
\usepackage{amssymb}	
\usepackage{float}
\usepackage{caption}
\usepackage{subcaption}
\usepackage{textcomp}
\usepackage{multirow}
\usepackage{makecell}
\usepackage{blindtext}
\usepackage{hyperref}
\usepackage{orcidlink}
\usepackage[normalem]{ulem}

\newcommand{\delSD}{$\Delta\Sigma_1$}
\newcommand{\lams}{$\lambda_{R_\mathrm{e}, \mathrm{Star}}$}

\newcommand{\lamg}{$\lambda_{R_\mathrm{e}, \mathrm{H_{\alpha}}}$}

\title[Kinematics vs. structure \& stellar population]{SDSS-IV MaNGA: Stellar rotational support in disk galaxies vs. central surface density and stellar population age}
\author[X. Wang et al.]{Xiaohan Wang\orcidlink{0009-0002-8312-1002}$^{1,2}$\thanks{E-mail:xiaohanw78@gmail.com},
Yifei Luo\orcidlink{0000-0001-7729-6629}$^{2}$\thanks{E-mail:yifeiluo@ucsc.edu},
S. M. Faber\orcidlink{0000-0003-4996-214X}$^{3}$,
David C. Koo$^{3}$,
Shude Mao\orcidlink{0000-0001-8317-2788}$^{1}$,
Kyle B. Westfall\orcidlink{0000-0003-1809-6920}$^{3}$,
\newauthor
Shengdong Lu\orcidlink{0000-0002-6726-9499}$^{4}$,
Weichen Wang$^{5}$,
Kevin Bundy$^{3}$,
N. Boardman$^{6}$,
Vladimir Avila-Reese$^{7}$,
\newauthor
Jos\'e G. Fern\'andez-Trincado$^{8}$,
Richard R. Lane\orcidlink{0000-0003-1805-0316}$^{9,10}$
\\
$^{1}$Department of Astronomy, Tsinghua University, Beijing 100084, China\\
$^{2}$Department of Astronomy and Astrophysics, University of California at Santa Cruz, Santa Cruz, CA 95064, USA\\
$^{3}$UCO/Lick Observatory, Department of Astronomy and Astrophysics, University of California, Santa Cruz, CA 95064, USA\\
$^{4}$Institute for Computational Cosmology, Department of Physics, University of Durham, South Road, Durham, DH1 3LE, UK\\
$^{5}$Universita degli Studi di Milano-Bicocca, Piazza della Scienza 3, 20126 Milano, Italy\\
$^{6}$School of Physics and Astronomy, University of St Andrews, North Haugh, St Andrews KY16 9SS, UK\\
$^{7}$Instituto de Astronomía, Universidad Nacional Autónoma de México, A.P. 70-264, 04510 México D. F., México\\
$^{8}$Instituto de Astronom\'ia, Universidad Cat\'olica del Norte, Av. Angamos 0610, Antofagasta, Chile\\
$^{9}$Centro de Investigaci\'on en Astronomi\'a, Facultad de Ingenieri\'a, Ciencia y Tecnologi\'a, Universidad Bernardo O'Higgins, \\Avenida Viel 1497, Santiago, Chile\\
$^{10}$Escuela de Ingenieri\'a Civil, Facultad de Ingenieri\'a, Ciencia y Tecnologi\'a, Universidad Bernardo O'Higgins, Avenida Viel 1497, \\Santiago, Chile\\
}

\date{Accepted XXX. Received YYY; in original form ZZZ}

\pubyear{2024}

\begin{document}
\label{firstpage}
\pagerange{\pageref{firstpage}--\pageref{lastpage}}
\maketitle

\begin{abstract}
We investigate how the stellar rotational support changes as a function of spatially resolved stellar population age ($\rm D_n4000$) and relative central stellar surface density (\delSD) for MaNGA isolated/central disk galaxies.
We find that the galaxy rotational support indicator $\lambda_{R_\mathrm{e}}$ varies smoothly as a function of \delSD\ and $\rm D_n4000$.
$\rm D_n4000$ vs. \delSD\ follows a ``J-shape'', with $\lambda_{R_\mathrm{e}}$ contributing to the scatters.
In this ``J-shaped'' pattern
rotational support increases with central $\rm D_n4000$ when \delSD\ is low but decreases with \delSD\ when \delSD\ is high. Restricting attention to low-\delSD\ (i.e, large-radius) galaxies, we suggest that the trend of increasing rotational support with $\rm D_n4000$ for these objects is produced by a mix of two different processes, a primary trend characterized by growth in $\lambda_{R_\mathrm{e}}$ along with mass through gas accretion, on top of which disturbance episodes are overlaid, which reduce rotational support and trigger increased star formation.
An additional finding is that star forming galaxies with low \delSD\ have relatively larger radii than galaxies with higher \delSD\ at fixed stellar mass.
Assuming that these relative radii rankings are preserved while galaxies are star forming then implies clear evolutionary paths in central $\rm D_n4000$ vs. \delSD. The paper closes with comments on the implications that these paths have for the evolution of pseudo-bulges vs. classical-bulges.
The utility of using $\rm D_n4000$-\delSD\ to study $\lambda_{R_\mathrm{e}}$ reinforces the notion that galaxy kinematics correlate both with structure and with stellar-population state, and indicates the importance of a multi-dimensional description for understanding bulge and galaxy evolution.

\end{abstract}



\section{Introduction}
\label{Introduction}
The rotational support of the stellar system in galaxies measures the relative magnitude of regular rotation to dispersion and shows important hints on galaxy evolution. The rotational support not only correlates with the galaxy specific angular momentum \citep{Cortese2016}, which is believed to determine the galaxy mass-size relation \citep{MMW1998,Shen2003}, but also observationally correlates with bulge-building processes, the transformation of mass profiles from exponential to de Vaucouleurs shape, and, ultimately, quenching \citep{Cappellari_2016,Graham2018}.

With the advent of integral-field spectroscopic (IFS) surveys \citep[e.g.][]{Bacon2001, deZeeuw2002, Cappellari2011,Croom2012, Sanchez2012,Bundy_2014}, an integrated parameter, $\lambda_{R_\mathrm{e}}$, has been invented to quantify rotational support and has proven to be powerful in characterizing galaxy spatial kinematics \citep[e.g.][]{2007MNRAS.379..401E,Cortese2016}.
To understand the connection between structure, stellar population and kinematics, the correlations of $\lambda_{R_\mathrm{e}}$ with various properties have been widely investigated. The first attempt focused on early-type galaxies and revealed their inhomogeneity in kinematic structures \citep[e.g.][]{2007MNRAS.379..401E, K2011, Cappellari_2016}.
Subsequent investigations expanded to a wider galaxy sample covering all morphological types. It was first shown with CALIFA galaxies \citep{Sanchez2012} that $\lambda_{R_\mathrm{e}}$ decreases with morphologies changing from spiral galaxies to elliptical galaxies \citep{Querejeta2015,  FB2019}. This morphology-kinematics correlation was further validated by SAMI \citep[Sydney-AAO Multi-object Integral field spectrograph,][]{Croom2012} \citep{Croom2021} and MaNGA \citep[Mapping Nearby Galaxies at Apache Observatory,][]{Bundy_2014} \citep{Graham2018}.

Morphology quantified by Hubble type is an approximate description of galaxy structure. The quantitative descriptions of galaxy structure, including Sérsic index, concentration ($R_{90}/R_{50}$), bulge-to-total ratio (B/T) and galaxy central density, help to characterize the galaxy mass distribution more accurately and reveal the complexity within each Hubble type.
The studies with quantitative structural parameters show consistent results, that $\lambda_{R_\mathrm{e}}$ tends to decrease with increasing concentration, Sérsic index and B/T \citep[e.g.][]{Querejeta2015,Cortese2016,FB2019, Croom2021, Cortese2022}.
The correlation of $\lambda_{R_\mathrm{e}}$ and galaxy mass distribution suggests that the change in $\lambda_{R_\mathrm{e}}$ is related to galaxy structural evolution, especially the growth of a dispersion-supported bulge. A recent study that performed kinematic bulge-disk decomposition for SAMI galaxies suggests that the $\lambda_{R_\mathrm{e}}$ of the galaxy can be fully explained by a combination of a dispersion-supported bulge and a rotation-supported disk \citep{Oh2020}.

Another possible factor that correlates with galaxy kinematics is galaxy stellar population. Observations of late-type galaxies have found that the disky structures tend to have younger stellar populations than bulges and stellar haloes
\citep{Lee2011,Sharma2014, GD2015, Goddard2017}. Therefore, the star formation history of a galaxy may influence the observed \lams\ values. Moreover, violent processes including mergers can dramatically change stellar orbits, decrease rotational support, and may lead to galaxy quenching \citep{Toomre1977,Hernquist1992,Naab2006}. A detailed view of how galaxy $\lambda_{R_\mathrm{e}}$ correlates with galaxy stellar age for a sample covering all morphological types is given by \citet{vandeSande2018}. They show that for both early-type and late-type galaxies, stellar population age tends to decrease with rotational support ($V/\sigma$, $\lambda_{R_\mathrm{e}}$) and galaxy intrinsic ellipticity. This correlation was further investigated by \citet{Wang2020} and  \citet{Fraser-McKelvie2021} who measured the distribution of stellar $\lambda_{R_\mathrm{e}}$ on and below the galaxy star formation main sequence (SFMS). They found that galaxies on the SFMS are always fast rotators, while below the SFMS galaxies there is a sharp decrease of $\lambda_{R_\mathrm{e}}$ at the high mass end, which they interpreted as being caused by multiple mergers for massive galaxies \citep{Wang2020}.

The two factors that correlate with kinematics, structural evolution and stellar population, are themselves correlated to some extent.
Under the hierarchical clustering model where galaxies grow by mergers, the bulge-to-total ratio (B/T) is expected to increase with stellar mass and stellar population age \citep[e.g.][]{White1978, Cole1994, Baugh2006}. This phenomenon has been seen in both observations and simulations \citep[e.g.][]{Kauffmann2003,Weinzirl2009, Hopkins2010, Bluck2014}.
Another scenario linking structural evolution and quenching is that a massive bulge can stabilize the galaxy and prevent further star formation \citep{Martig2009}.
Therefore, a general correlation among structure, kinematics and stellar population is expected. The observations consistently show that, in a general way, all of the structural, stellar population and kinematic properties correlate with one another and form a sequence. One end of the sequence is characterized by high random motions, older stellar populations, and higher bulge-to-total ratios. The other end of the sequence is characterized by high rotational support, active star formation, younger stellar populations, and lower bulge-to-total ratios \citep[e.g.][]{KK04,Fisher2006, FD2008, Gadotti2009, Fisher_Drory2010, Fabricius_2012, FD2016, FB2019, Croom2021}.
An illustration of the sequence can be seen in \citet{Croom2021}, which shows a simultaneous decline in \lams\ with increasing concentration and declining specific star formation rate for non-elliptical galaxies.
A theoretical interpretation summarizing the kinematic and structural evolution for galaxies on and below the SFMS can be found in \citet{Wang2020}.

However, all of the correlations along the sequence show considerable scatter.
The inconsistency can be seen from high values of Sérsic index and B/T on the SFMS \citep{Wuyts2011,Morselli2017}, indicating that star-forming galaxies can have highly concentrated mass distributions.
The inconsistency was further investigated under the effort to find the best structural predictor of galaxy quenching, which uncovers that structures and stellar populations evolve at different rates as galaxies begin to quench.
\citet{Cheung_2012} introduced the quantity $\Sigma_1$, which is the projected stellar density within 1 kpc radius. They showed that $\Sigma_1$ was the best predictor of quenching among a variety of structural parameters measured for galaxies near $z \sim 1$. \citet{Fang_2013} defined the quantity \delSD\ by subtracting the mass trend from $\Sigma_1$ and plotted \delSD\ vs. global $NUV-r$ for a large sample of SDSS galaxies. The resulting figures showed an ``elbow shape'' in which low \delSD\ galaxies were uniformly star-forming whereas high \delSD\ galaxies could be either star-forming or quenched (illustrated further in Fig. \ref{classification}). Finally, \citet{Barro_2017} and \citet{Lee2018} showed that the elbow pattern is present as early as $z \sim 3$, and thus this feature has been a fundamental property of galaxy evolution since very early times.

The relationship of structure to star formation was explored further by \citet{Luo2020} using various structural and stellar parameters for a large sample of SDSS galaxies. They reproduced the elbow patterns with \delSD\ and central $\rm D_n4000$ (the 4000-\AA \ break strength, stellar population age indicator). \citet{Luo2020} also compared \delSD\ with \citet{Gadotti2009}'s bulge classification parameter $\Delta \mu_e$ from the Kormendy relation. The two parameters were found to agree well, making \delSD\ a good indicator of bulge structure.
In summary, all of these works revealed the same trend in which quenched galaxies always have high central densities while star-forming galaxies can have a range of central surface densities. Said differently, low-relative-central-density galaxies are always star-forming while high-relative-central-density galaxies can have a range of star formation rates.

Adding kinematics to this picture adds additional scatter.
On the one hand, galaxies below the SFMS consistently show high B/T or Sérsic indices \citep[e.g.][]{Wuyts2011,Morselli2017}, but significantly different \lams\ at the low/high mass end \citep[e.g.][]{Wang2020}. This difference was interpreted as resulting from different quenching mechanisms at low and high mass.
\citet{Cortese2022} used $\Sigma_1$ to characterize galaxy concentration for purely passive galaxies but found no correlation between $\Sigma_1$ and $\lambda_{R_\mathrm{e}}$ for this sample at fixed mass.
On the other hand, although galaxies on the SFMS are typically fast rotators, the scatter of $\lambda_{R_\mathrm{e}}$ is always considerable, even among star forming galaxies \citep{vandeSande2018,FB2019, Wang2020, Fraser-McKelvie2021}.

To summarize, galaxy stellar rotational support ($\lambda_{R_\mathrm{e}})$ correlates broadly with structural and stellar population properties, but superimposed on those general relations are additional subtrends that seem to be controlled by other variables.
This complexity is hardly surprising, as it has been clear for many years that galaxies populate a multi-dimensional manifold, with the result that virtually every simple scaling law in two dimensions shows residuals that typically correlate with other parameters.
With that as background, it is therefore natural to look at $\lambda_{R_\mathrm{e}}$ in the same way, i.e., to map its behavior in a higher-dimensional space that spans both structure and stellar-population state.
That exploration in detail will be the topic of a future paper, and here we take a first step by mapping $\lambda_{R_\mathrm{e}}$ in just two dimensions, using \delSD\ to encapsulate structural information and central/global values of $\rm D_n4000$ to describe the stellar population.
Because elliptical galaxies are heavily influenced by mergers \citep{Springel2005b, Kormendy2009, Naab2009} and satellite galaxies may be influenced by environmental factors \citep{Peng2010}, we focus on disk galaxies and central/isolated galaxies only in this paper.
We use kinematic data from MaNGA (Mapping Nearby Galaxies at Apache Point Observatory) \citep{Bundy_2014} to study how the rotational support parameter $\lambda_{R_\mathrm{e}}$ varies across the 2-D landscape of $\rm D_n4000$ vs. \delSD.
The choice of $\rm D_n4000$ and \delSD\ as coordinates proves fortunate: 
a new ``J-shape'' pattern emerges whereby $\lambda_{R_\mathrm{e}}$ increases with central $\rm D_n4000$ when \delSD\ is low but decreases with \delSD\ when \delSD\ is high. The trends for low-\delSD\ objects are especially clean and suggest a possible evolutionary path for these objects in star formation rate and $\lambda_{R_\mathrm{e}}$ vs. stellar mass. Still other correlations, for kinematically and morphologically disturbed galaxies, are described along the way.

The paper is organized as follows. Section \ref{Data_and_Analysis} describes the data and sample selection, introduces the parameters we computed, and discusses possible biases in the measurements.
Section \ref{disk_kinematics} presents our main finding of the J-shaped pattern in central $\rm D_n4000$ vs. \delSD. 
In Section \ref{pseudobulge_M_Dn4000} we show how galaxy rotational support depends on stellar mass and central stellar population age for low-\delSD\ galaxies and offer a simple evolutionary scenario to account for the observed trends.  In Section \ref{Discussion} we discuss possible evolutionary tracks on $\rm D_n4000_{center}$ vs. \delSD\ space and re-express our findings for $\lambda_{R_\mathrm{e}}$ in the language of pseudo-bulges and classical bulges.
Section \ref{Conclusions} summarizes our results and open issues.
In this paper we adopt a concordance $\Lambda$CDM cosmology: $H_0 = 70 \mathrm{km/(s\;Mpc)}$, $\Omega_\mathrm{M} = 0.3$ and $\Omega_{\Lambda} = 0.7$.

\section{DATA AND ANALYSIS}
\label{Data_and_Analysis}

\subsection{MaNGA IFU observations}
The MaNGA (Mapping Nearby Galaxies at Apache Point Observatory) Survey provides spatially resolved spectra and property maps for $\sim 10000$ galaxies in the local universe ($z < 0.15$) \citep{Bundy_2014}. It utilizes the 2.5m Sloan Foundation Telescope in its spectroscopic mode \citep{Gunn_2006} and the two dual channel BOSS spectrographs \citep{Smee_2013}, which cover a wavelength range of $3600-10300$\AA\;with R $\sim 2000$ and spatial resolution of $\sim 2.5\arcsec$ \citep{Bundy_2014}. The instrumental details are described in \citet{Drory_2015}.
Galaxies in this paper are selected from MaNGA Project Launch-11 (MPL-11), the latest and most complete version of MaNGA.
Data used in this paper are from the SPX MAPS provided by the MaNGA Data Analysis Pipeline \citep[DAP, ][]{Westfall_2019, Belfiore2019}, which makes use of pPXF \citep{Cappellari_2004, Cappellari_2017} to process spectra and produce 2-D maps of properties.

\subsection{Photometric and spectroscopic measurements}

For sample selection and kinematic measurements, we adopt elliptical Petrosian effective radii from the extended NASA-Sloan Atlas (NSA) catalog \citep{Blanton2011, Wake_2017}, galaxy axis ratio $b/a$ from \citet{Simard_2011}'s catalog, and stellar mass from MPA-JHU DR7 value-added catalog \citep{2003MNRAS.341...33K}.
The choice of elliptical Petrosian $R_\mathrm{e}$ is consistent with MaNGA target selection \citep{Wake_2017}, as NSA's single Sérsic $R_\mathrm{e}$ tends to be systematically overestimated for galaxies with higher Sérsic indices \citep{Simard_2011, Wake_2017}.

We also adopt the smoothness parameter $S2$ from \citet{Simard_2011}.
The smoothness parameter $S2$ is defined in \citet{Simard2009}, calculated as a sum of fractional total light and asymmetry light of image residuals from a smooth symmetric model, and therefore is an indicator of galaxy asymmetry and clumpiness \citep{Schade1995, Simard_2002}. A higher $S2$ indicates higher asymmetry or higher clumpiness. We adopt $S2_g$, which is calculated with the g-band SDSS images. The choice of band for $S2$ does not have a large effect on our results.

\begin{figure}
    \centering
    \includegraphics[width = \columnwidth]{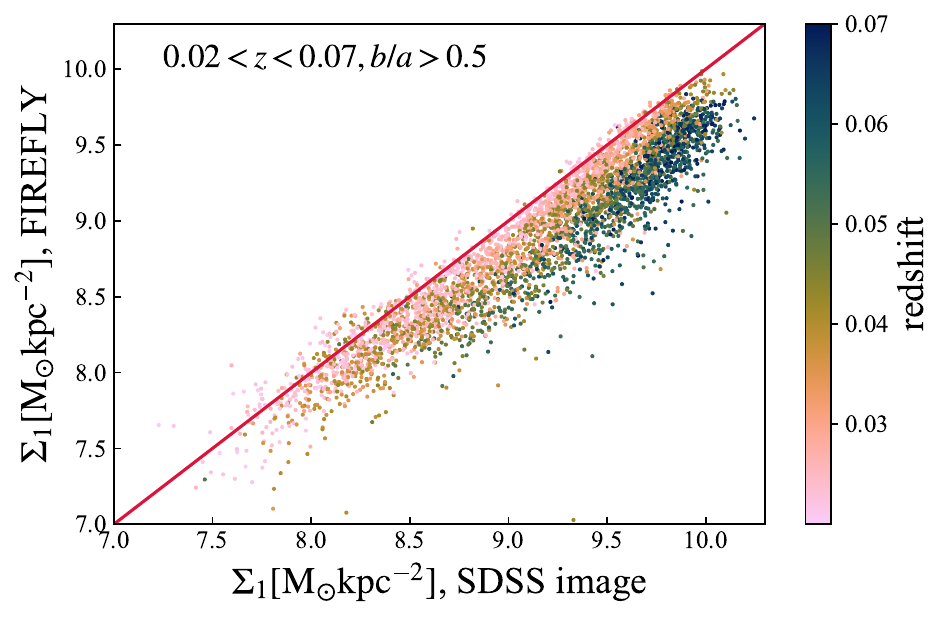}
    \caption{$\Sigma_1$ calculated by MaNGA mass maps generated by FIREFLY vs. $\Sigma_1$ calculated by color difference in SDSS images, colored by redshift. The sample is limited to $0.02 < z < 0.07$, $b/a > 0.5$ (See sample selection in Sec. \ref{sample_selection}).
    The $\Sigma_{1, \rm MaNGA}$ is systematically lower than $\Sigma_{1, \rm SDSS}$, and the difference increases with redshift, consistent with the spatial resolution difference of MaNGA and SDSS.
    We use SDSS measurements in our analysis.}
    \label{2SD1}
\end{figure}

The central density is quantified by \delSD, 
the residual of $\Sigma_1$ to stellar mass, as defined in \citet{Luo2020}. $\Sigma_1$, the stellar surface mass density within the radius of 1 kpc, is measured with i-band luminosities of SDSS DR7 images \citep{Abazajian_2009} and i-band mass-to-light ratios from \citet{Fang_2013}. \delSD\ is calculated by a function which separates the two density peaks in $\Sigma_1$ at fixed stellar mass 
\citep{Luo2020}:
\begin{equation}
    \Delta \Sigma_1 = \rm \log \Sigma_1 + 0.275(\log M_*)^2 -6.445\log M_* + 28.059.
\end{equation}

The spatial resolution for SDSS ($\sim 1.5\arcsec$) is better than MaNGA ($\sim 2.5\arcsec$), and the luminosity profiles of SDSS images are seeing-corrected for $\Sigma_1$ measurements (Zhao et al. in preparation). Therefore, we choose to adopt $\Sigma_1$ calculated from SDSS DR7 images instead of recalculating it with data from MaNGA.
Fig. \ref{2SD1} compares $\Sigma_1$ from SDSS images with $\Sigma_1$ calculated by the mass maps provided by MaNGA FIREFLY value-added catalog \citep{Neumann2022}, colored by redshift.
As expected by the difference of spatial resolution, $\Sigma_{1, \mathrm{FIREFLY}}$ is systematically smaller than $\Sigma_{1, \mathrm{SDSS}}$, and the offset increases with redshift.

The stellar populations of galaxies are characterized by the 4000-\AA \ break strength, $\rm D_n4000$.
In this paper, we calculate mean $\mathrm{D_n4000}$ in radii of $1.5 \arcsec$ (SDSS fiber size) and bundle size (MaNGA field of view) from MaNGA maps, weighted by SPECINDEX\_WGT \citep[flux of linear continuum, see Sec. 5.2.2 in][]{A2022, Westfall_2019} provided in MaNGA DAP.
Fiber measurements are denoted as ``center'', while bundle measurements are denoted as ``global''. These labels represent that these measurements serve as indicators of the central and the overall stellar population ages. The choice of the bundle size ensures that the global measurements are effectively averaged across the entire galaxy.
We checked the results by using $\mathrm{D_n4000}_{R_\mathrm{e}}$ as indicators of global stellar population ages and found that the main conclusions are not affected.

\subsection{Galaxy Sample}
\label{sample_selection}

\begin{figure}
    \centering
    \includegraphics[width=\columnwidth]{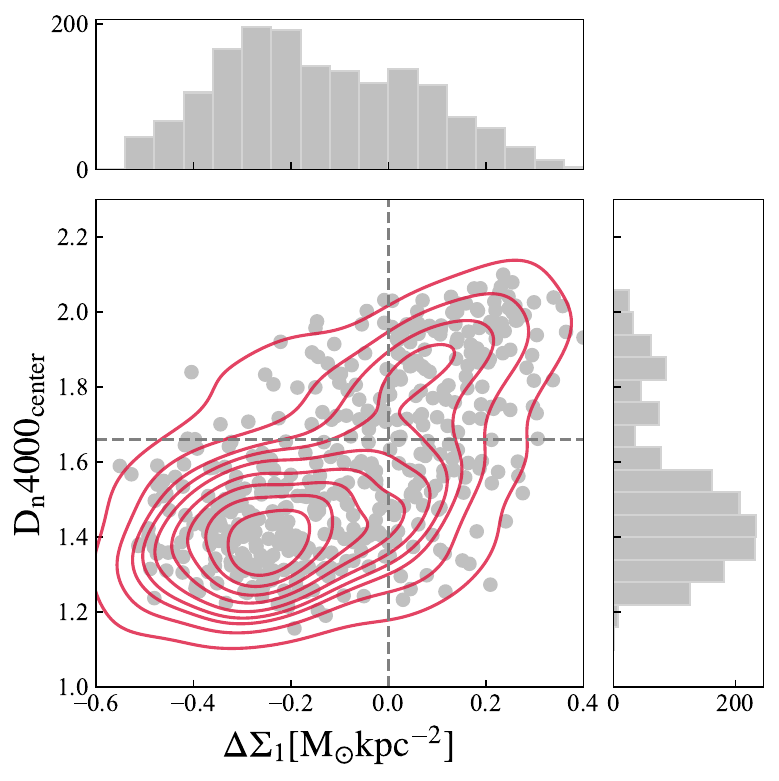}
    \caption{$\mathrm{D_n4000_{center}}$ vs. \delSD\ and their histograms, showing how the ``elbow pattern'' studied by \citep{Fang_2013} and \citep{Luo2020} is revealed in this sample. The subscript ``center'' means the parameter is measured within a $1.5\arcsec$ radius to match the SDSS fiber, same in the following text. The crimson curves represent weighted number density contours, where the points are weighted by \texttt{esweight} divided by the sampling fraction (see text), to correct the sample into a volume-limited sample.
    The vertical dashed line \delSD\ $= 0 $ is the classification boundary of bulge types in \citet{Luo2020}.
    The horizontal gray dashed line $\mathrm{D_n4000_{center}} = 1.66$ is a rough indicator of green valley.
    In this way, galaxies are divided into four bins, centrally star-forming galaxies with high \delSD\ (lower right), centrally star forming galaxies with low \delSD\ (lower left), centrally quiescent galaxies with high \delSD\ (upper right) and centrally quiescent galaxies with low \delSD\ (upper left).}
    \label{classification}
\end{figure}

\begin{figure*}
    \centering
    \includegraphics[width = \textwidth]{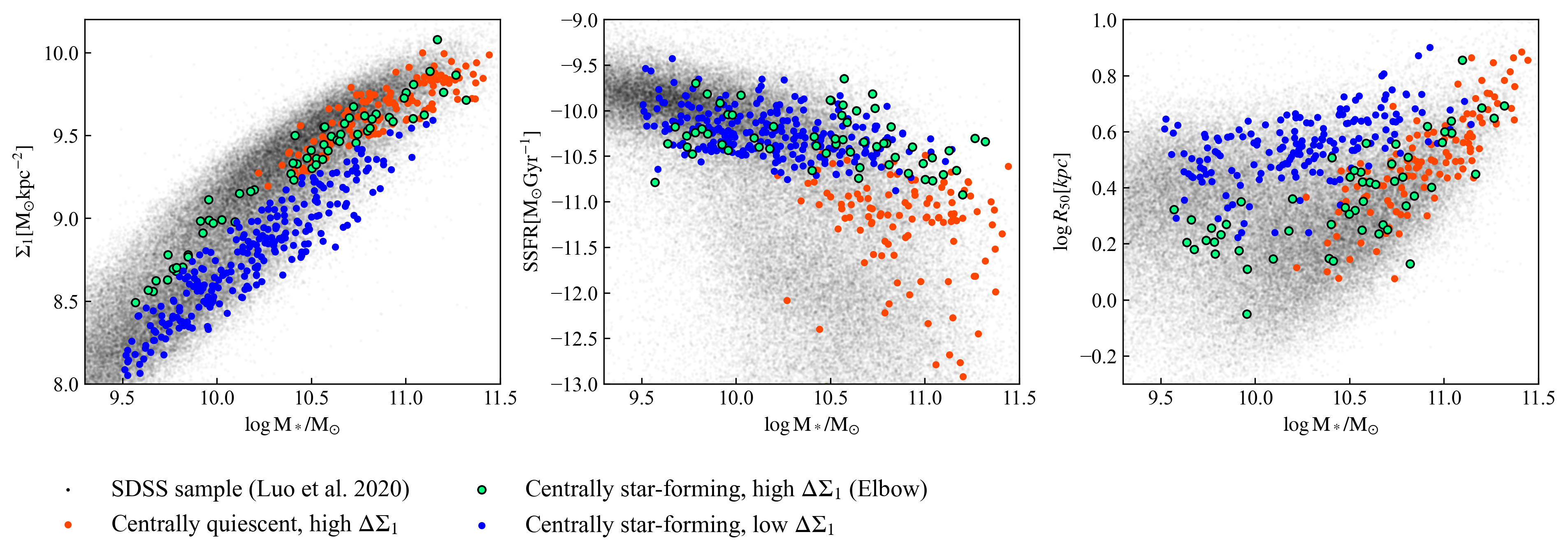}
    \caption{From left to right are $\Sigma_1$ vs. stellar mass, specific star formation rate vs. stellar mass and half-mass radius vs. stellar mass. The gray points in the background are SDSS sample with $0.02 < z < 0.07$ and $b/a > 0.5$. The red points are centrally quiescent high-\delSD\ galaxies (\delSD\ $>0$, $\mathrm{D_n4000_{center}} > 1.66$) in our sample, the green open circles are elbow galaxies (\delSD\ $>0$, $\mathrm{D_n4000_{center}} < 1.66$) and the blue points are centrally-star-forming low-\delSD\ galaxies (\delSD\ $<0$, $\mathrm{D_n4000_{center}} < 1.66$).
    The left two figures represent the definitions of the three subsets. Specifically, low-\delSD\ galaxies have lower central density and tend to be star-forming, while high-\delSD\ galaxies have higher central density and cover a large range of star formation.
    As shown in the third figure, low-\delSD\ galaxies always have larger radii than high-\delSD\ galaxies at fixed mass. This size difference is consistent with the central densities. This size difference also sheds light on bulge evolution, which is discussed in Sec. \ref{Discussion}.
    }
    \label{sample}
\end{figure*}

\subsubsection{Main Sample selection}
We reiterate here that our goal is to study well resolved galaxies with substantial disk components. With that in mind, we set the following requirements, which are summarized in Table \ref{table_selection}:
\begin{enumerate}
    \item Galaxies are limited to $0.02 < z < 0.07$.
    
    Nearby galaxies ($z < 0.02$) are rejected as their intrinsic proper motions can affect distance measurements. Galaxies are restricted to $z < 0.07$ to mitigate the seeing effect for $\Sigma_1$, as when $z > 0.07$, the apparent size of 1 kpc will be smaller than $0.7 \arcsec$, which is half of seeing for SDSS images \citep{Fang_2013, Luo2020}. 

    \item Galaxies are limited to have stellar mass higher than $10^{9.5} \rm M_{\odot}$.
    
    In \citet{Luo2020}, only galaxies with mass higher than $10^{9.5} \rm M_{\odot}$ are fitted to obtain the \delSD\ function.

    \item Only isolated or central disk galaxies are included. 
    
	Satellite galaxies can be affected by their environmental effects. In this paper we only focus on isolated or central galaxies. Satellite galaxies are excluded by group designation ($M_\mathrm{rank} > 1$) from \citet{Yang2012}.

    The morphology classification is from Galaxy Zoo 1 \citep{Lintott2011} and we exclude galaxies with the value of \texttt{ELLIPTICAL} = 1. However, the visual classification may not work well and may result in losing disk galaxies. We checked the classification for the whole MaNGA sample and found some of the galaxies with \texttt{ELLIPTICAL} = 1 show low Sérsic indices, which are possible lost disk galaxies. Nevertheless, the galaxies with \texttt{ELLIPTICAL} = 1 and Sérsic indices < 2 cover only a small fraction (0.7\%) compared to galaxies with \texttt{ELLIPTICAL} = 0. Therefore, we simply adopt the classification based on \texttt{ELLIPTICAL} = 1.

    \item The probability of merging ($P_{\mathrm{MG}}$, from Galaxy Zoo 1 \citep{Lintott2011}) is lower than 0.2.

    \item Axis ratio $b/a$ is from 0.5 to 0.85.
    
    The axis ratio is required to be larger than $0.5$ to exclude near edge-on galaxies, where $\Sigma_1$ can be overestimated by disk component contamination or reduced by dust obscuration.
    The axis ratio is required to be smaller than $0.85$ to reject nearly face-on galaxies whose inclination corrections are not reliable. We make inclination correction for velocity, but measurements of $b/a$ for nearly face-on galaxies may be not accurate and the error of correction can be severe. We made a test with a limited sample of $ 0.5 < b/a < 0.7$ and the main results do not change.
    In addition, the $b/a$ measurements, adopted for inclination correction, may not well represent the true inclination angle due to non-zero disk height and can be biased by structures like bars and asymmetric spiral arms. We checked the sample by eye and rejected all edge-on galaxies and apparently face-on galaxies remaining after the $b/a$ cut.
    
    \item The effective radius is larger than $5\arcsec$, and the galaxy is covered to at least $1.2R_\mathrm{e}$ by the MaNGA bundle.
    
	Beam smearing effects can cause $\lambda_{R_\mathrm{e}}$ to be underestimated. By requiring $R_\mathrm{e} > 5\arcsec$, we select galaxies sampled by at least 4 beams along their major axes, such that their kinematics are better resolved.
    Simulations, discussed in Appendix \ref{spatial_resolution}, indicate that beam-smearing effects on $\lambda_{R_\mathrm{e}}$ are small for galaxies larger than $5\arcsec$.
	We also require galaxies to be covered to $1.2 R_\mathrm{e}$ to obtain accurate and reliable measurements within $R_\mathrm{e}$. This allows us to avoid spaxels near the edge of the IFU field-of-view, which may suffer from low SNR and unreliable kinematics.

    \item No galaxies with ``bad quality''.
    
    Spaxels flagged by the MaNGA DRP as DONOTUSE or FORESTAR, or with SNR < 3 are masked in our analysis.
    The flag of DONOTUSE, however, varies with properties for the same spaxel, and we make our selection based on kinematic measurements.
    We exclude galaxies whose fraction of rejected spaxels within 1 $R_\mathrm{e}$ ellipse is larger than 10\%.

    \item The misalignment between the photometry and kinematic major axes must be $ < 30\degr$.
    
    Due to the effect of bars, the major axis of photometry may be misaligned with the kinematic major axis. As we adopt photometric $R_\mathrm{e}$ and $b/a$ for kinematic measurements, we calculate kinematic major axis position angles for stellar velocity fields using \texttt{fit\_kinematic\_pa} \footnote{http://purl.org/cappellari/software}, and exclude galaxies whose misalignment between photometry and kinematic major axes is larger than $30\degr$.
\end{enumerate}

The above cuts yield a sample of 484 galaxies. The resulting sample sizes under each criterion are summarized in Table \ref{table_selection}.

The distribution of the sample on $\mathrm{D_n4000_{center}}-$\delSD\ is shown in Fig. \ref{classification}.
The distribution shows the classic elbow shape, where galaxies with low \delSD\ are always strongly star-forming but galaxies with high \delSD\ cover a large range of star formation rate. 
The distributions in both x and y-axis are bimodal. Their minima, \delSD\ $= 0 $ and $\mathrm{D_n4000_{center}} = 1.66$ (a rough ridgeline for green valley), can be used to define four quadrants for future reference:
centrally star-forming galaxies with high \delSD, centrally star forming galaxies with low \delSD, centrally quiescent galaxies with high \delSD\ and centrally quiescent galaxies with low \delSD. The last type only contains few galaxies and is not discussed in this paper.
This classification is consistent with \citet{Luo2020} who uses \delSD\ $=0$ as a classification boundary.

Fig. \ref{sample} shows the distributions of the three main types of galaxies on $\Sigma_1-\rm M_*$, $\mathrm{SSFR}-\rm M_*$ and $R_\mathrm{50}-\rm M_*$, with background of SDSS sample \citep{Luo2020}.
An important feature is that galaxies with low \delSD\ always have larger radii than galaxies with high \delSD\ at a fixed mass. This feature will be further discussed in Sec. \ref{Discussion}.

The data points in Fig. \ref{classification} are weighted to correct the sample to be volume-limited.
We adopt a method like that of \citet{Fraser-McKelvie_2022}, by calculating the ``sampling rate'' of our sample, which is the fraction of selected galaxies to a MaNGA parent sample ($0.02< z < 0.07$, $\rm \log M_*/M_\odot > 9.5$, \texttt{esweight} $>0$) in bins of $\mathrm{M_*}-z$. We then weight each galaxy by \texttt{esweight} \citep[provided by MaNGA, see][]{Wake_2017} divided by the number fraction. 
This correction is applied for the density plot (Fig. \ref{classification}) and when calculating correlation coefficients (Figs.  \ref{PB_subsets}, \ref{struc}). We clarify that the correction cannot fully avoid the bias introduced by the requirement of effective radii ($R_{\rm e} > 5\arcsec$), and the sample is biased to larger galaxies. Discussion about this bias is in Sec. \ref{Caveats} and will be further addressed in a future work (Wang et al. in prep.).

\begin{table}
    \centering
    \caption{Sample selection descriptions.}
    \begin{tabular}{lcr}
        \hline
        & Criteria & N \\
        \hline
        & $0.02 < z < 0.07$ & 7735\ \\
        & $0.5 < b/a < 0.85$ & 3270 \\
        & Isolated or Central & 2170 \\
        & Disk Galaxies & 1671 \\
        & Merging Probability $ < 0.2$ & 1589 \\
        & $\rm M_* > 10^{9.5} M_{\odot}$ & 1347 \\
        & $R_\mathrm{e} > 5\arcsec$, IFU size $> 1.2R_\mathrm{e}$ & 666 \\
        & \texttt{esweight} $ > 0$ & 649 \\
        & Checked by Eye for Good Inclination & 584 \\ 
        & Good Data Quality & 533 \\
        & \makecell{No misalignment between \\ photometry and kinematics} & 484 \\
        \hline
    \end{tabular}
    \label{table_selection}
\end{table}

\subsubsection{Clean Sample and Subsamples}
\label{Subsamples}

Galaxy kinematics may be affected by disturbances. External disturbances, including mergers, interactions and gas infall, may result in asymmetric or irregular morphologies and alter stellar orbits. Internal disturbances, often associated with bars, can also change the orbits and may heat the stars.
To investigate the effects of disturbance on galaxy kinematics, we define three subsets, ``disturbed morphology'', ``perturbed kinematics'', and ``clean sample'', from the main sample.

To select the galaxies with disturbed morphologies, we set $S2_g > 0.17$, which is the median value of the main sample, as a preliminary selection, and further update the selection by checking galaxy images.
We flag the galaxies with relatively irregular or asymmetric morphologies with ``disturbed morphology'', which are possible to have experienced mergers or interactions. Examples of such objects are shown in Fig. \ref{PB_gallery}.
Then we check the velocity maps for all galaxies in our sample, and flag the galaxies with twisted central velocity maps as ``perturbed kinematics''.
Most often such disturbances are associated with bars.
Notice that the two subsets of ``disturbed morphology'' and ``perturbed kinematics'' are not mutually exclusive; they have an intersection of 13 galaxies.
Finally, based on these analyses, we define the clean sample as those galaxies not flagged as having either ``disturbed morphology'' or ``perturbed kinematics''. 
We return to these peculiar subsets again in Sec. \ref{pseudobulge_M_Dn4000}.
The criteria and numbers of each subset are shown in Table \ref{sub_samples}.

\begin{table}
    \centering
    \caption{Description of subsamples.}
    \begin{tabular}{lccr}
        \hline
        & Description & Criteria & N \\
        \hline
        & \makecell{Disturbed \\Morphology} & \makecell{irregular or asymmetric \\morphology} & 33 \\
        \hline 
        & \makecell{Perturbed \\ Kinematics} & \makecell{centrally twisted velocity maps}& 232 \\ 
        \hline 
        & Clean Sample & \makecell{ Settled disks, \\regular morphology \\and kinematics}& 232 \\
        \hline
    \end{tabular}
    \label{sub_samples}
\end{table}

\subsection{Proxy for rotation support: $\lambda_{R_\mathrm{e}}$}
\label{kin_lam_vsigma}

\begin{figure*}
    \centering
    \includegraphics[width=\textwidth]{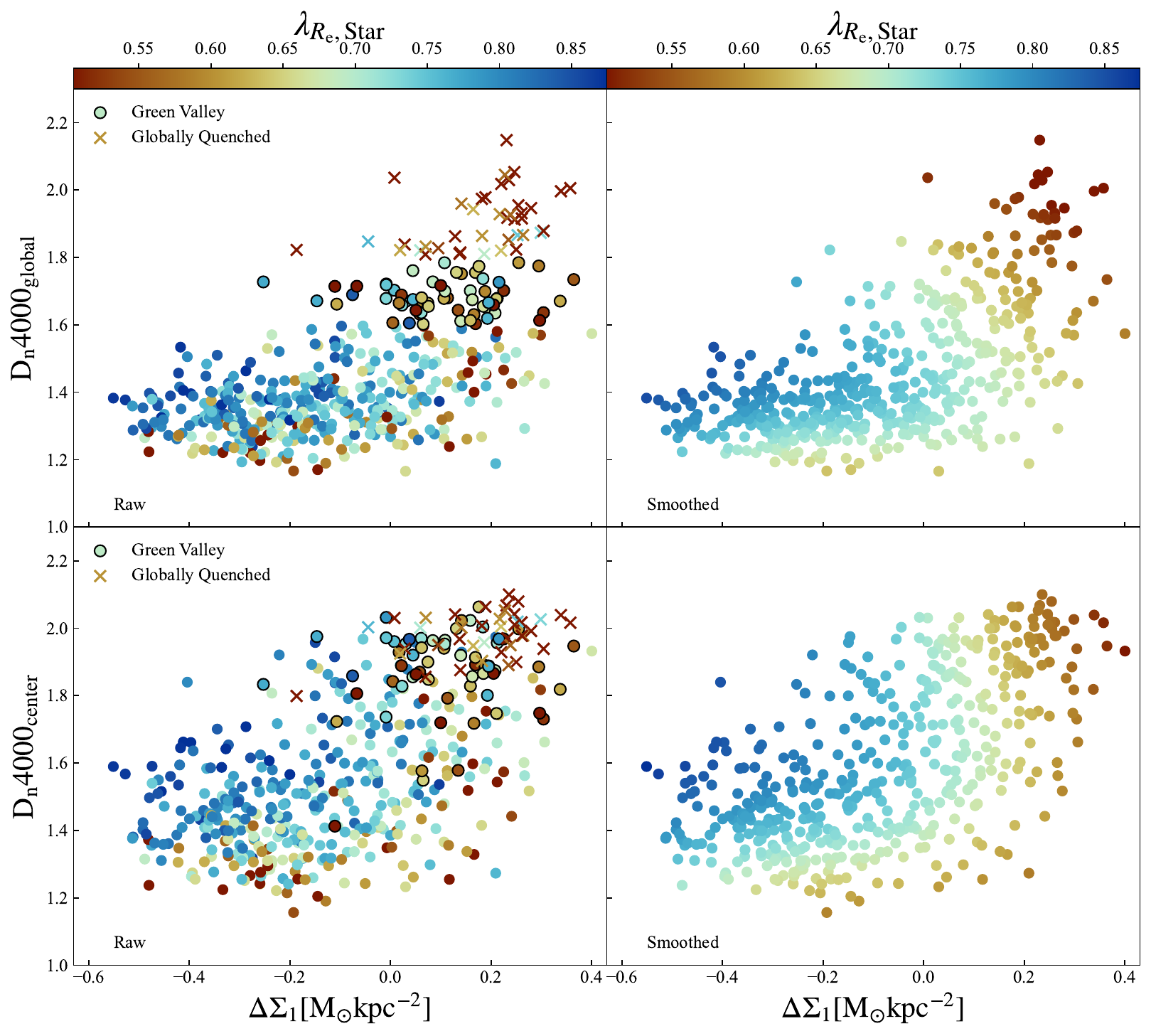}
    \caption{$\mathrm{D_n4000_\mathrm{global}}$ and $\mathrm{D_n4000_{center}}$ vs. \delSD\ colored by \lams. The left column shows the original scatter plots and the right panels are LOESS smoothed plots. Fully quenched galaxies ($\mathrm{D_n4000_\mathrm{global}} > 1.8$) are labeled as cross markers and galaxies in green valley ($1.6 < \mathrm{D_n4000_\mathrm{global}} < 1.8$) are circles with black edges.
    The distribution of \lams\ on $\mathrm{D_n4000_\mathrm{global}}$ vs. \delSD\ space shows a non-monotonic tilted pattern, where \lams\ first increases and then decreases with $\rm D_n4000_{global}$.
    The distribution of $\lambda_{R_\mathrm{e}}$ on central $\rm D_n4000$ vs. \delSD\ is ``J-shaped'', in which $\lambda_{R_\mathrm{e}}$ increases as $\rm D_n4000_{center}$ increases for when \delSD\ is low, and decreases as \delSD\ increases when \delSD\ is high.}
    \label{Dn4000_center}
\end{figure*}

We use $\lambda_{R_\mathrm{e}}$ as the galaxy rotational support indicator. 
$\lambda_{R_\mathrm{e}}$ is a dimensionless proxy for observed projected specific angular momentum, defined in \citet{2007MNRAS.379..401E}. The definition of $\lambda_{R_\mathrm{e}}$ via two-dimensional spectroscopy is
\begin{equation}
    \lambda_R = \frac{\sum_{i=1}^{N}F_i R_i |V_i|}{\sum_{i=1}^{N}F_i R_i\sqrt{V_i^2 + \sigma_i^2}},
\end{equation}
where $F_i$, $R_i$, $V_i$ and $\sigma_i$ are the flux, projected circular radius, velocity and velocity dispersion of the $i\mathrm{th}$ spatial bin \citep{2007MNRAS.379..401E}.
In the original definitions of $\lambda_{R_\mathrm{e}}$ in \citet{2007MNRAS.379..401E}, $V_i$ and $\sigma_i$ represent projected line-of-sight kinematics. However, it is possible to make proper inclination correction for velocity of disk galaxies. 
We correct $V$ to be $V/\sin i$, and assume that $\sigma$ is isotropic and does not need correction. Therefore, kinematic measurements in this paper are \emph{deprojected}.
The parameter $\sin i$ is calculated by:
\begin{equation}
    \sin^2{i}  = \frac{1-(b/a)^2}{1-\alpha^2},
\end{equation}
where $\alpha$ is the ratio of disk scale height over disk scale length and $b/a$ is the axis ratio. This formula is based on the oblate model and from \citet{1958MeLuS.136....1H}. The value of $\alpha$ is observed to range from 0.1 to 0.3 for local galaxies \citep{Pandilla2008, Unterborn2008, R2013}. We adopt a typical value of 0.2 and apply it to all galaxies in our sample.
$\lambda_{R_\mathrm{e}}$ is calculated in ellipses with kinematic major axis, photometric axis ratio and Petrosian $R_\mathrm{e}$.

$\lambda_{R_\mathrm{e}}$ may be underestimated due to the beam smearing effect.
Empirical corrections for beam smearing effect on $\lambda_{R_\mathrm{e}}$ have been developed \citep[]{Graham2018, Harborne2020}. However, the correction method is tested with observation data for early-type galaxies only \citep{Graham2018}, which are not covered in this work.
Therefore, we choose to not apply any beam smearing correction for $\lambda_{R_\mathrm{e}}$ in this paper. To estimate the effect of beam smearing effect, we have made a simulation, which is similar to the method in \citet{Greene2018}. We have found that, for galaxies with $R_\mathrm{e}$ larger than $5\arcsec$, the beam smearing effect is not significant and does not bias our results. Details of the simulation are in Appendix \ref{spatial_resolution}.
The beam smearing correction will be explored for a wider range of galaxies in future work (Wang et al. in prep.).

In our analysis, dispersions are corrected for instrumental resolution with $\sigma_\mathrm{corr}$ provided by the MaNGA DAP.
However, a problem arises for regions with low dispersion, where errors in observations can result in imaginary values of the corrected dispersion.
Ignoring such pixels can lead to biased overestimation of dispersion values.
It is recommended in \citet{Westfall_2019} to retain these negative $\sigma^2$ values, and we keep these values and utilize a luminosity-weighted mean squared dispersion to account for their contributions.
This approach ensures a more accurate representation of the dispersion properties in our analysis.

\section{Rotational support as a function of relative central density and stellar population}
\label{disk_kinematics}

\begin{figure*}
    \centering
    \includegraphics[width=\textwidth]{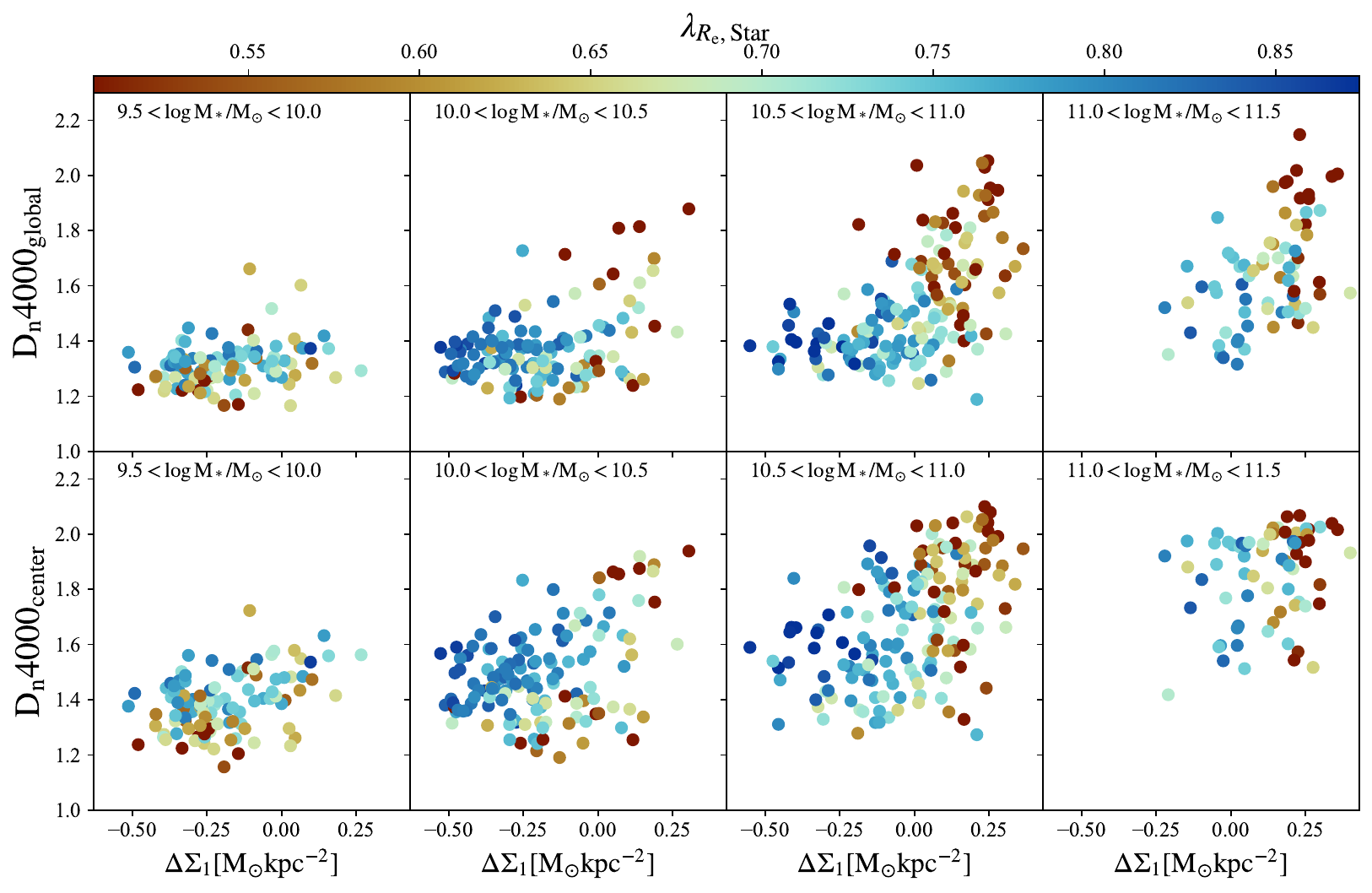}
    \caption{\lams\ as a function of $\rm D_n4000$ and \delSD\ in four mass bins. The top and bottom panels separately show $\mathrm{D_n4000_\mathrm{global}}$ and $\mathrm{D_n4000_{center}}$ vs. \delSD\ colored by \lams. As stellar mass increases, the whole population moves from lower left to higher right along the elbow-shaped distribution. The color pattern gradually changes from horizontal to vertical, indicating the main dependence of kinematics changes from stellar population to structures (see text).}
    \label{Dn4000_global}
\end{figure*}

In this section we present how galaxy rotational support, indicated by $\lambda_{R_\mathrm{e}}$, varies as a function of stellar population ($\rm D_n4000$) and relative central density (\delSD).

Fig. \ref{Dn4000_center} shows the distribution of \lams\ in the $\mathrm{D_n4000_\mathrm{global}}$-\delSD\ and $\mathrm{D_n4000_\mathrm{center}}$-\delSD\ spaces.
Colors in the right panels are smoothed by the LOESS method\footnote{http://purl.org/cappellari/software} \citep{Cappellari2013b}, an implementation of two-dimensional Locally Weighted Regression \citep{Cleveland_Devlin_1988}, to get a view of the average distribution.
We mark globally quenched galaxies ($\mathrm{D_n4000_\mathrm{global}} > 1.8$) as crosses and galaxies in the green valley ($1.6 < \mathrm{D_n4000_\mathrm{global}} < 1.8$) as edged circles, to better trace their locations when central $\rm D_n4000$ is used.

The most intriguing feature in Fig. \ref{Dn4000_center} is the smoothly varying color pattern in both global and central parameter spaces.
For global $\rm D_n4000$-\delSD\ (top row), the distribution of \lams\ follows tilted stripes, where \lams\ increases first with global $\rm D_n4000$ when star forming, and then decreases as galaxies enter the green valley and quench. The stripes are almost horizontal when \delSD\ is negative, where \lams\ is weakly dependent on \delSD, and get tilted where \lams\ decreases with both \delSD\ and global $\rm D_n4000$ when \delSD\ is high.
This pattern of horizontal and tilted stripes can be described as a ``fainted J-shape'', where \lams\ increases with global $\rm D_n4000$ when \delSD\ is low, and decreases with both \delSD\ and global $\rm D_n4000$ when \delSD\ is high.
The ``J-shape'' pattern is more prominent for central $\rm D_n4000$ -\delSD\ in the bottom row, where the color stripes are almost horizontal for negative \delSD\ and get vertical for positive \delSD.
\lams\ increases with central $\rm D_n4000$ when \delSD\ is low, and decreases with \delSD\ when \delSD\ is high.

For both global and central parameter spaces, the ``dominant'' parameter that correlates best with \lams\ changes with the locations in the space, with a transition at almost \delSD\ $\sim 0$.
When \delSD\ is high, \lams\ tends to decrease with both $\rm D_n4000_{global}$ and \delSD\ with a tilted pattern, but shows almost no correlation with central $\rm D_n4000$.
This is possibly owing to the intermixing of the green valley and globally quenched objects when central $\rm D_n4000$ is used.
In retrospect, the complexity at high \delSD\ is not unexpected, as galaxies with high \delSD\ show a wide range of star formation rates \citep[e.g.][]{Fang_2013,Luo2020}. 
By contrast, galaxies with low \delSD\ are more homogeneous with a narrower range of $\rm D_n4000$, and are found to have similar behavior in both parameter spaces, where \lams\ tends to increase with both central and global $\rm D_n4000$, with only a quite weak dependence on \delSD.
This trend is opposite to the general trend mentioned in the Introduction that \lams\ tends to decrease with increasing stellar population age.
In particular, \lams\ shows a tighter correlation with central $\rm D_n4000$ than global $\rm D_n4000$. This is shown by the fact that the low-\delSD\ galaxies with different \lams\ are closely mixed in global $\rm D_n4000$ but more widely spread out in central $\rm D_n4000$.
In other words, central $\rm D_n4000$ is a better predictor for \lams, which is measured within 1 $R_\mathrm{e}$.
This correlation between central stellar population and global kinematics, that is opposite to the general relation, will be further investigated in the next section.

It is natural to ask whether the above patterns are actually mass-effect patterns.
To investigate this, we divide the sample into four mass bins and show their distributions separately in Fig. \ref{Dn4000_global}.
The first impression from Fig. \ref{Dn4000_global} is the bulk movement of populations in the diagram with stellar mass.
The galaxy population slowly moves from lower left to upper right, with \delSD\ and $\rm D_n4000$ both increasing, as stellar mass increases.
This is consistent with the general trend that \lams\ tends to decrease with $\rm D_n4000$ and \delSD, as the data move from the horizontal arm (dominated by high \lams) to the vertical arm (dominated by low \lams) of the elbow distribution. 
Therefore, stellar mass does play a role in determining the shape of the overall distribution and the general relations.
However, the color distributions are not simply caused by mass effect, as shown by the fact that the elbow distribution is present separately in the two middle mass bins and by the fact that the horizontal and vertical color patterns are each individually visible over a a wide range of mass.

The new ``J-shaped'' pattern reproduces the familiar trends that \lams\ tends to decrease with $\rm D_n4000$ and \delSD, but reveals a more complex pattern underlying both trends. 
In addition to the large-scale trends, it has also been found that stellar rotational support increases first and then decreases with stellar population age \citep{FB2019,Croom2021}, consistent with what is shown by the J-shape pattern.
Furthermore, investigations on \lams\ and galaxy structure found that \lams\ has a weak dependence on $R_{90}/R_{50}$ with a relatively flat slope when $R_{90}/R_{50}$ is low \citep{FB2019,Croom2021}, consistent with the relatively horizontal color patterns in Fig. \ref{Dn4000_center} when \delSD\ is low.
Investigations utilizing \delSD\ show that \lams\ decreases with \delSD\ for star-forming galaxies but no trends for passive galaxies \citep{Cortese2022}. This finding also agrees with the color distribution in each mass bin, as shown in the top panel of Fig. \ref{Dn4000_global}, which further specifies that introducing central/global stellar population can help reduce the scatter of \lams\ vs. \delSD\ for star-forming galaxies.
In summary, the J-shape pattern is consistent with previous observations, and in particular uncovers the whole pattern and subtrends by utilizing a 2-D coordinate that combines stellar population and structure.

\section{Possible kinematic evolution of disk galaxies with low relative central surface density}
\label{pseudobulge_M_Dn4000}

\begin{figure}
    \centering
    \begin{subfigure}{\columnwidth}
        \includegraphics[width = \columnwidth]{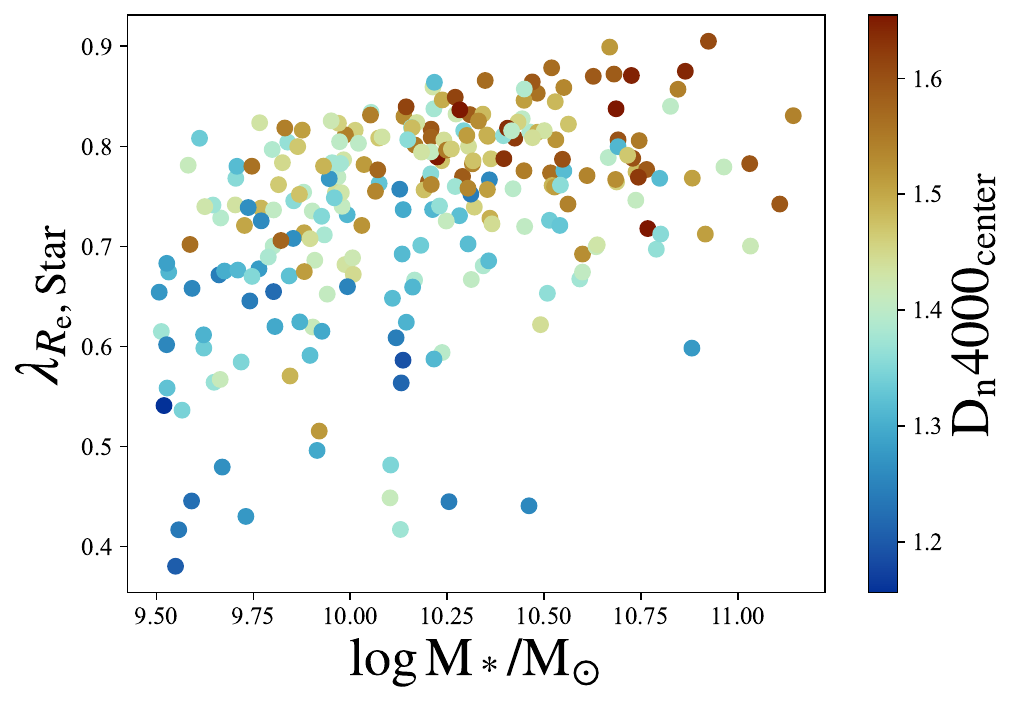}
        \caption{\lams\ vs. stellar mass colored by central $\rm D_n4000$.}
        \label{lambda_m_Dn4000}
    \end{subfigure}
    \begin{subfigure}{\columnwidth}
        \includegraphics[width = \columnwidth]{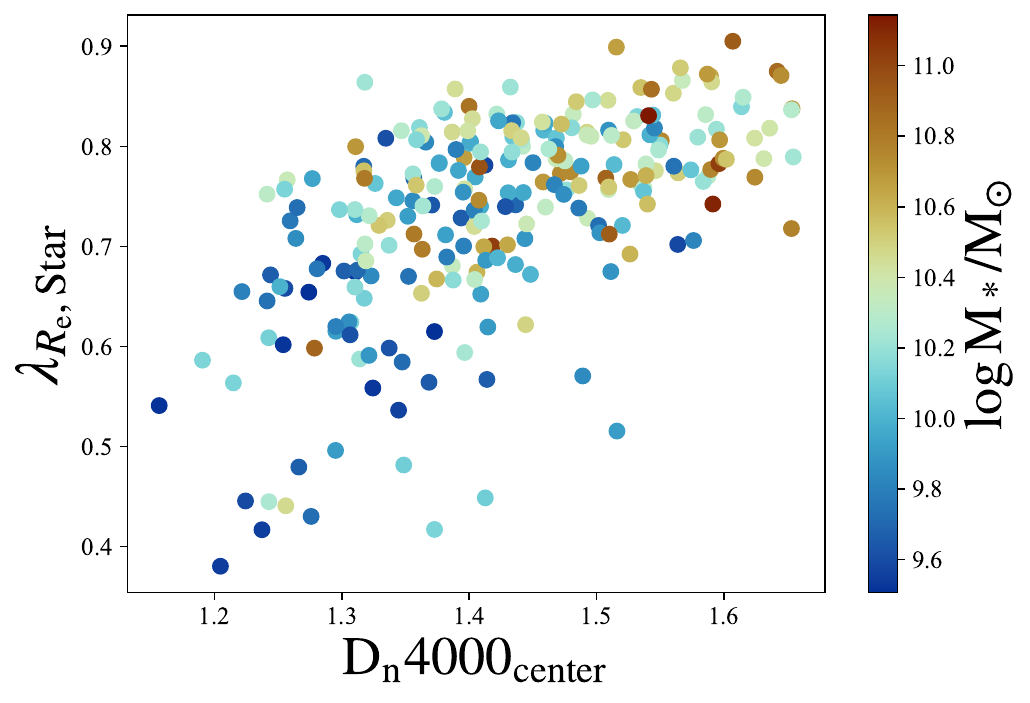}
        \caption{\lams\ vs. central $\rm D_n4000$ colored by stellar mass.}
        \label{lambda_Dn4000_m}
    \end{subfigure}
    \caption{The correlations among \lams\, stellar mass, and central $\rm D_n4000$ for galaxies with \delSD$<0$, $\rm D_n4000_{center} < 1.66$.}
    \label{lam_Dn4000_m}
\end{figure}

\begin{figure*} 
    \centering
    \includegraphics[width = \textwidth]{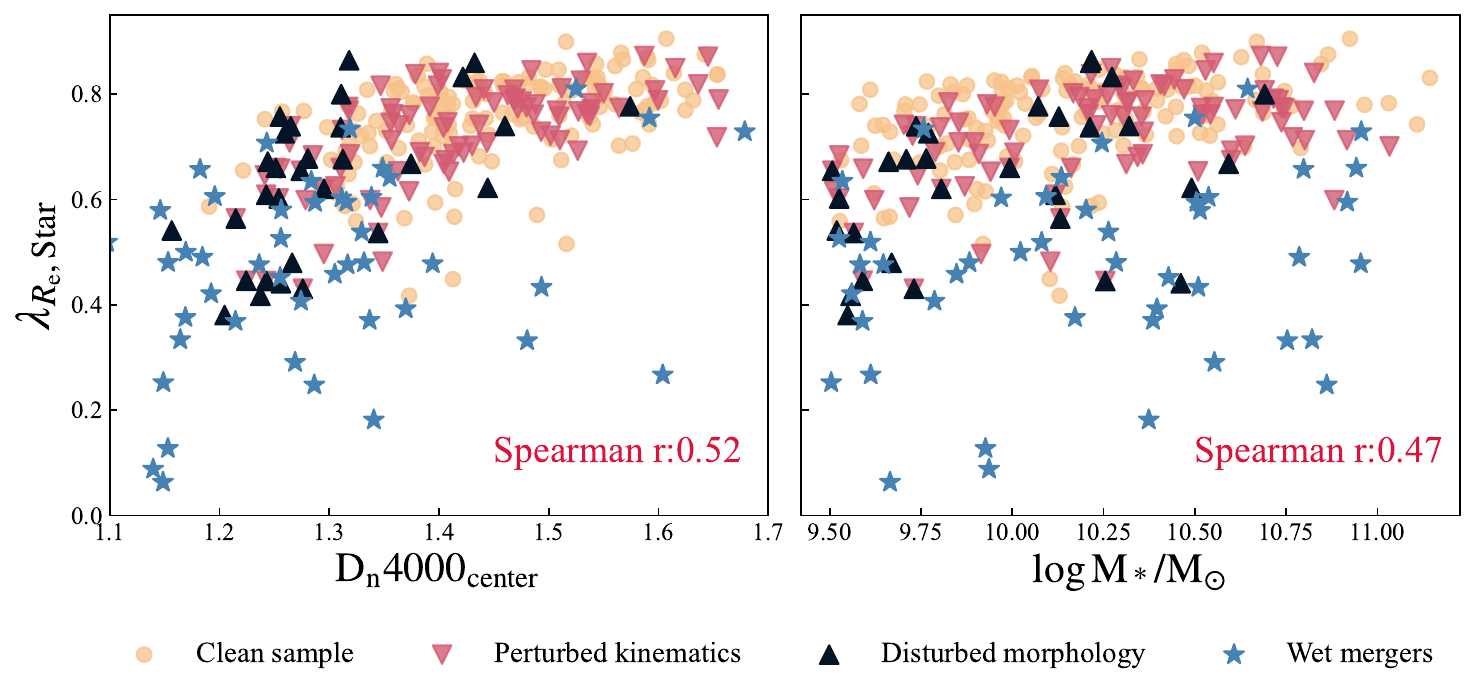}
    \caption{Replots of Fig. \ref{lam_Dn4000_m} with kinematically/morphologically disturbed galaxies tagged and star-forming merging galaxies added.
    Only galaxies with \delSD\ < 0, $\mathrm{D_n4000_{center}} < 1.66$ are shown. Galaxies are flagged with ``clean sample'', ``perturbed kinematics'', and ``disturbed morphology'' as introduced in Sec. \ref{sample_selection}.
    Star-forming merging galaxies are shown by the blue stars. They have $P_{\rm MG} > 0.3$ and $\rm SSFR > -10.5$, and are termed ``wet mergers'' here on account of their high star-formation rates.
    They are not part of the main sample and were excluded in plotting Fig. \ref{lam_Dn4000_m} and computing Spearman correlation coefficients.
    It is seen that \lams\ increases more with $\mathrm{D_n4000_{center}}$ than stellar mass (see Spearman correlation coefficients), and that morphologically disturbed galaxies and merging galaxies cluster at low $\rm D_n4000_{center}$. We speculate in the text that the distribution is consistent with a mixed evolutionary scenario, in which disturbances, which tend to reduce \lams\ temporarily, are superimposed on an underlying evolutionary trend in which $\lambda_{R_{\rm e}}$ and $\rm D_n4000$ increase with mass (see text).}
    \label{PB_subsets}
\end{figure*}

\begin{figure*} 
    \centering
    \begin{subfigure}{\textwidth}
        \includegraphics[width = \textwidth]{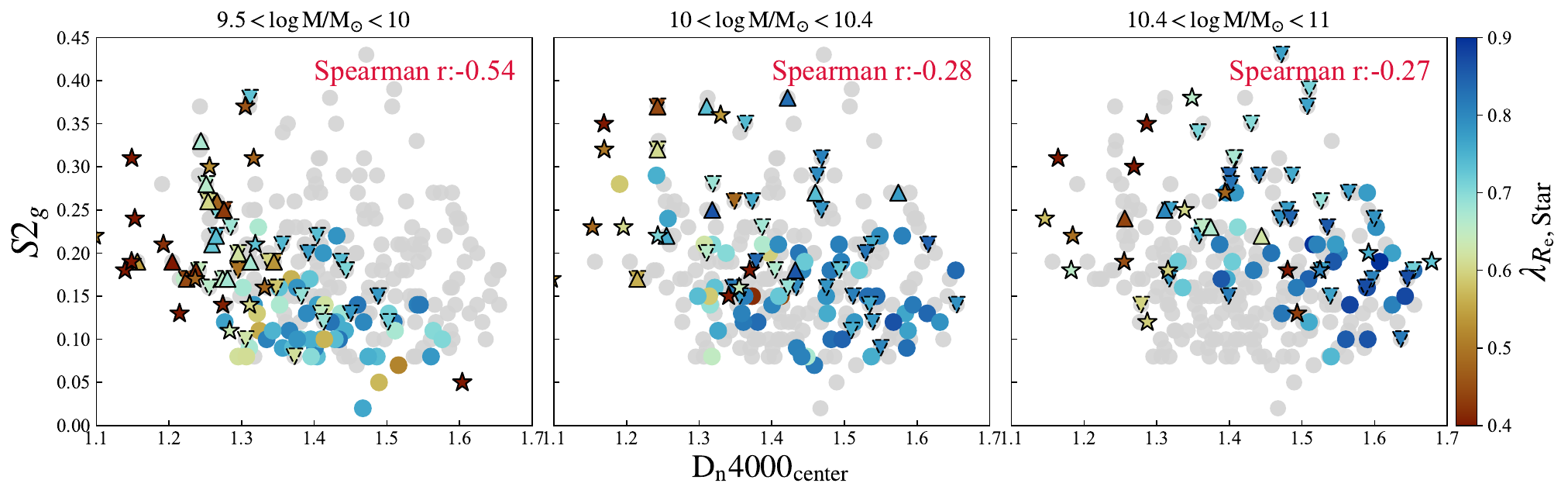}
    \end{subfigure}
    \begin{subfigure}{\textwidth}
        \includegraphics[width = \textwidth]{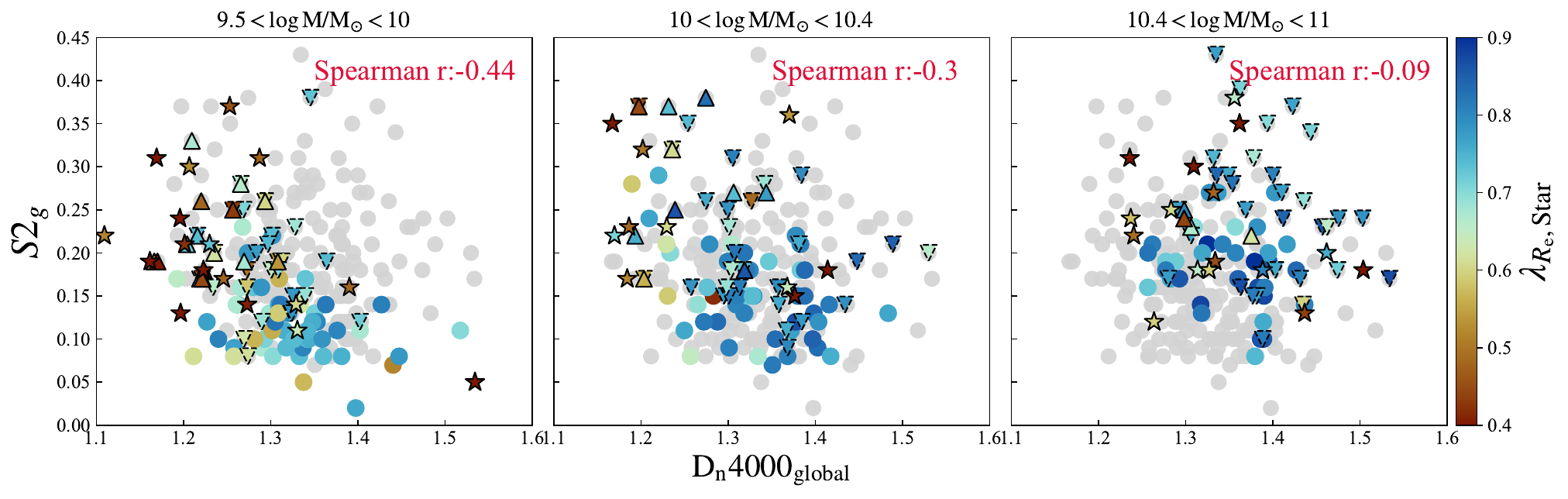}
    \end{subfigure}
    \begin{subfigure}{\textwidth}
        \includegraphics[width = \textwidth]{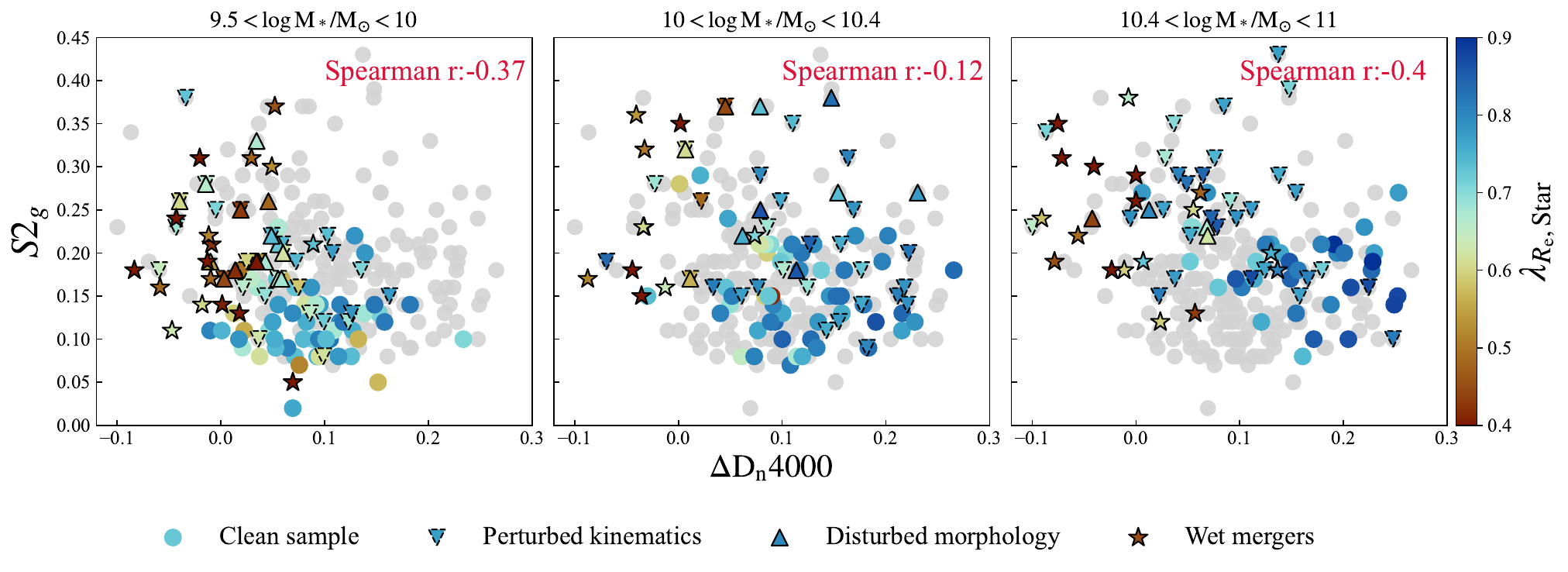}
    \end{subfigure}
    \caption{$S2_g$ vs. $\mathrm{D_n4000_{center}}$, $\rm D_n4000_{center}-D_n4000_{global}$, $\rm D_n4000_{global}$, colored by \lams\ in three mass bins for star-forming low-\delSD\ galaxies (\delSD\ < 0, $\mathrm{D_n4000_{center}} < 1.66$) and merging galaxies ($P_{\rm MG} > 0.3, \rm SSFR > -10.5$). The gray points on the background are the whole sample of galaxies with \delSD\ < 0, $\mathrm{D_n4000_{center}} < 1.66$.
    The merging galaxies are shown as stars, and are excluded when computing Spearman correlation coeﬃcients.
    From left to right the colored points are $\rm 9.5 < \log M_*/M_{\odot} < 10$, $\rm 9.5 < \log M_*/M_{\odot} < 10.4$, and $\rm 10.4 < \log M_*/M_{\odot} < 11$. In each mass bin, $S2_g$ decreases both with $\rm D_n4000_{center}$ and $\rm \Delta D_n4000$, with \lams\ increasing. Similar trends are also seen in low and middle mass bins for $\rm D_n4000_{global}$. The behavior of $S2_g$, \lams\ and $\rm D_n4000$ in each mass bin is consistent with the scenario that galaxies are disturbed and undergo central star formation.}
    \label{struc}
\end{figure*}

In this section we focus on galaxies with low \delSD, whose \lams\ typically increase with $\rm D_n4000$, particularly for central stellar populations.
This trend exists within each mass bin (the left three bins in Fig. \ref{Dn4000_global}), while \lams\ also tends to increase with stellar mass on average, shown as the increasing fraction of deep blue points versus mass for low-\delSD\ galaxies in Fig. \ref{Dn4000_global}. To get a better view of how \lams, $\rm D_n4000_{center}$ and mass correlate with each other, we select the galaxies with negative \delSD\ and $\rm D_n4000_{center} < 1.66$ (252 galaxies), where the trend of \lams\ versus $\rm D_n4000$ is the clearest, and plot their distribution on \lams\ vs. mass, coloring the points by central $\rm D_n4000$.
The results are shown in Fig. \ref{lambda_m_Dn4000}.
A complementary view is shown in Fig. \ref{lambda_Dn4000_m} of \lams\ vs. central $\rm D_n4000$ with the points colored by stellar mass.

Fig. \ref{lam_Dn4000_m} clearly indicates the aforementioned trends of increasing \lams\ with stellar mass and $\rm D_n4000$, and increasing \lams\ with central $\rm D_n4000$ at a fixed stellar mass. A new feature revealed in Fig. \ref{lam_Dn4000_m} is that both distributions consist of a relatively tight ridgeline, along which \lams\ increases with $\rm D_n4000_{center}$ and stellar mass, accompanied by a fringe population of low \lams\ ``outliers'' lying below.

The relations of \lams\ vs. stellar mass and stellar population have been investigated in previous studies. Overall trends are that \lams\ decreases with stellar mass \citep[e.g.][]{Brough2017,vandeSande2017,Veale2017,Greene2018} and stellar population age \citep[e.g.][]{vandeSande2018,FB2019,Wang2020,Fraser-McKelvie2021}, but a closer reading reveals higher-order complexities. For example, a decrease in \lams\ at low stellar mass is visible in some samples \citep{FB2019,vandeSande2021a,vandeSande2021b}, as well as a decrease in \lams\ for highly star-forming galaxies, also at the low mass end \citep{Wang2020, Croom2021, Fraser-McKelvie2021}.
The results of Figs. \ref{Dn4000_center}-\ref{lam_Dn4000_m} together now show that these individual subtrends can be understood as parts of a larger, more complicated, but smoothly varying higher-dimensional manifold.
The general trend masks the existence of different and possibly opposite subtrends \emph{within} the population, which we are dissecting here by isolating the sub population.

We now turn to the question of the formation channels of the ridgeline and the scattered outliers in Fig. \ref{lam_Dn4000_m}.
Since massive galaxies are further along the arc of star-formation vs. mass and have older stellar populations, the increasing trend of rotational support versus stellar mass may also contribute to the upward slope of \lams\ vs. $\rm D_n4000_{center}$ in Fig. \ref{lambda_Dn4000_m}. That is, $\rm D_n4000_{center}$ and rotational support may be evolving together with increasing stellar mass.
This simple scenario is consistent with the increasing specific angular momentum of the accreted gas with time \citep{Kassin2012,Simons2017,Renzini2020}, which makes galaxies spin-up in disks as they assemble their mass by forming stars \citep{Simons2017,Peng2020,Renzini2020}. Similar processes are also visible in simulations \citep{Lagos2017}.
Late-accreted gas with higher specific angular momentum can hardly reach the center and feed the star formation there, leading to central stellar population aging with increasing mass \citep{Renzini2018}.
This scenario, however, cannot explain the low-lying outliers or the strong tendency of low \lams\ galaxies at a fixed mass to have low $\rm D_n4000_{center}$. Additional ideas are needed to explain these features.

As described in Sec. \ref{sample_selection}, we divide galaxies into subsets interpreted as affected by different kinds of disturbance.
Galaxies flagged as having ``perturbed kinematics'' show twist signatures in their inner velocity fields that plausibly originate from non-axisymmetric, \emph{internal} mass irregularities such as bars.
Galaxies denoted as having ``disturbed morphologies'' exhibit asymmetric or clumpy structures in their images, which are more likely caused by \emph{external} interactions, such as a merger or a fly-by or perhaps simply the recent arrival of a large amount of gas in one coherent blob.
To further estimate the effect of external disturbances, we now also call back the star-forming merging galaxies which were rejected when choosing the sample (Sec. \ref{sample_selection}).
All three samples -- disturbed morphologies, perturbed kinematics, and merging galaxies -- are plotted in \lams\ vs. $\rm D_n4000_{center}$ and \lams\ vs. $\rm \log M_*$ in Fig. \ref{PB_subsets}.
The rejected merging galaxies (limited to have $\rm P_{MG} > 0.3, \log M_*/M_{\odot} > 9.5$) are shown as the blue stars, which are further limited to objects with $\rm SSFR > -10.5$ in order to ensure that all galaxies be comparably gas-rich.

Fig. \ref{PB_subsets} shows several trends.
The two subsamples, perturbed-kinematics and disturbed-morphology, cover most of the low $\lambda_{R_\mathrm{e}}$ outliers, hence disturbance seems to be associated with low apparent rotational support.
Galaxies with ``perturbed kinematics'' (pink triangles) are more dispersed however, not only occupying low $\lambda_{R_{\rm e}}$ regions but also residing on the ridge line of clean sample. This suggests that internal perturbations, like bars, are less serious kinematic disturbances and do not inevitably imply low rotational support.
In contrast, galaxies with ``disturbed morphologies'' (black triangles) mainly reside in low $\lambda_{R_{\rm e}}$ and low $\rm D_n4000_{center}$ regions, with a relatively small fraction reaching the ridgeline. Low $\lambda_{R_{\rm e}}$ and young $\rm D_n4000_{center}$ values correlate most tightly with disturbed morphology, indicating that external disturbances usually depress \lams\ and central $\rm D_n4000$.
This interpretation is borne out by the distribution of clearly merging galaxies (blue stars), which lie lowest in the diagrams of all samples. Their departure from the ridgeline is in the same direction but even stronger than the disturbed galaxies.
We suggest that these objects, together with the black triangles, exhibit a continuum of external interactions, with the blue stars representing the most extreme examples. These points are also illustrated by direct images in Fig. \ref{PB_gallery} which shows a montage of galaxies arranged by central $\rm D_n4000$.

\begin{figure}
    \centering
    \includegraphics[width = 0.48\textwidth]{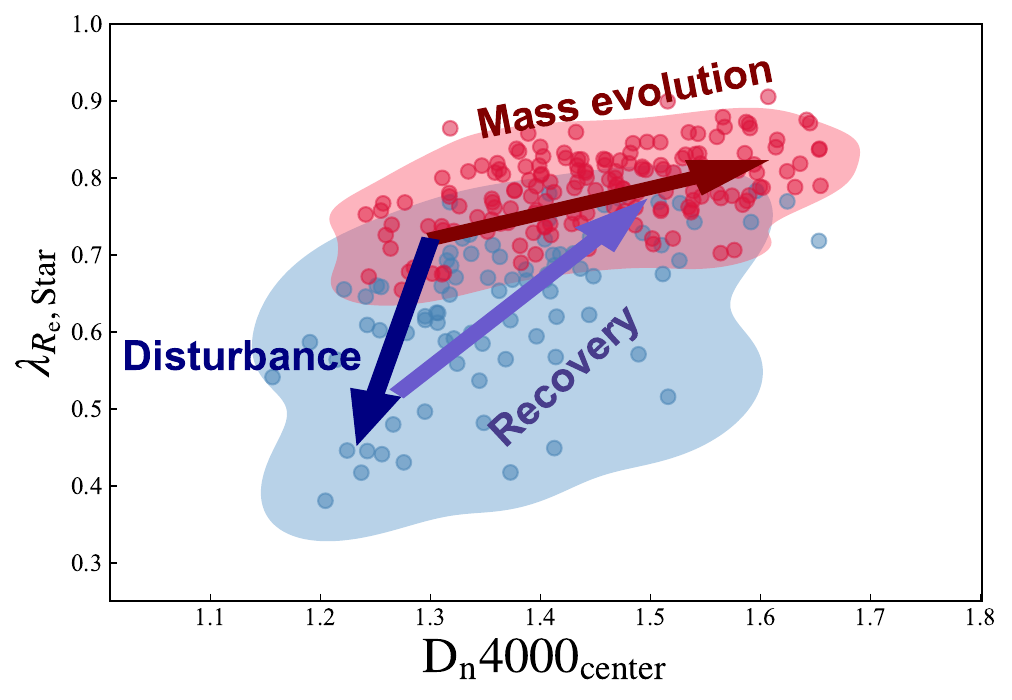}
    \caption{Schematic summary of possible evolutionary paths under a mixed-scenario model for galaxies with negative \delSD\ and $\rm D_n4000_{center} < 1.66$ in \lams\ vs. $\rm D_n4000_{center}$ diagram. The red points are defined as top sequence in the \lams\ vs. $\rm M_*$ plot. The red arrow shows the evolution revealed by stellar mass, whereby rotational support and central age increase.
    The dark blue arrow illustrates how disturbances might effectively drag a galaxy to low \lams\ and $\rm D_n4000_{center}$ values. The disturbed galaxies can gradually recover spin by forming new stars from late-accreting, high-angular momentum gas, thereby getting back to the sequence along the light purple arrow.}
    \label{disturbance_diagram}
\end{figure}

We further test these inferences in by introducing $S2$, a quantitative indicator of galaxy clumpiness and asymmetry, as an indicator of external disturbance.
The clumpiness parameter $S2$, defined by \citet{Simard2009}, responds to all types of asymmetric structures, including bars and spiral arms, and is therefore not quite ideal for spotting mergers and major disturbances. However, the former features are more prevalent at high masses, and at the lower masses considered here, $S2$ is useful. 
With $S2$ as a disturbance indicator, the disturbance evolutionary scenario predicts decreasing $S2$ with increasing central $\rm D_n4000$, and the trends should be seen at a fixed mass. In particular, the trend could be stronger for lower mass, as low-mass galaxies are more easily disturbed.
Finally, this trend should be present for global $\rm D_n4000$ as well as central $\rm D_n4000$, as external disturbances may enhance star formation globally as well as centrally.

Fig. \ref{struc} tests these predictions by plotting g-band $S2$ vs. central/global $\rm D_n4000$ and also $\rm D_n4000$ gradient in three mass bins, and coloring the points by \lams.
The merging galaxies are also included in Fig. \ref{struc}, shown as stars and colored by \lams\ as well.
The results are generally consistent with these predictions.
$S2_g$ is seen to decrease with increasing $\rm D_n4000$ within each mass bin, with \lams\ increasing along the trend, as expected.
The trend of decreasing $S2_g$ is the strongest in the lowest mass bin, consistent with the scenario that low mass galaxies are more easily disturbed.
The similar trends are also visible for $\rm D_n4000$ gradient in all mass bins, and $\rm D_n4000_{global}$ for the left mass bins, consistent with the prediction that disturbance would enhance global star formation and flatten the radial star formation gradients.
The only exception is for the highest masses, where $\rm D_n4000_{global}$ shows almost no correlation with $S2_g$, but $\rm D_n4000_{center}$ and $\Delta \rm D_n4000$ do.

In summary, it would appear that at least two separate processes acting in concert are needed to explain all the features of low-\delSD\ galaxies as shown in Figs. \ref{lam_Dn4000_m}-\ref{struc}. 
A hypothetical ``mixed scenario'' to do this is illustrated in Fig. \ref{disturbance_diagram}, which schematically shows how low-\delSD\ galaxies might actually move in \lams\ vs. $\rm D_n4000_{center}$ under this scenario.
The red points in Fig. \ref{disturbance_diagram} are selected to be on the ridgeline of \lams\ vs. stellar mass ($\lambda_{R_\mathrm{e}} > 0.1\log \rm M_*/M_\odot -0.3$). This selection reproduces the sequence in \lams\ vs. $\rm D_n4000$, as shown in Fig. \ref{disturbance_diagram}. The blue points, by contrast, are low-$\lambda_{R_\mathrm{e}}$ outliers in \lams\ vs. $\rm \log M_*$. The three arrows summarize the different processes that affect galaxy \lams\ and star formation state.
The foundational trend is the tendency of \lams\ and $\rm D_n4000_{center}$ to increase as galaxies gain mass while star-forming and settle into dynamically colder, older disks, which in turn depress star formation in the center. Superimposed on this main trend are occasional perturbations caused either by arrivals of major packets of new gas or possibly by outright mergers, dragging galaxies off the main mass-evolution ridgeline (blue arrow). These can affect galaxies of any mass but are slightly more frequent at low mass, as indicated by Fig. \ref{PB_subsets}. Their tendency to generate excess central star formation makes them cluster strongly at low values of $\rm D_n4000_{center}$.
Finally, the disturbed galaxies may gradually recover their kinematics and return to the high \lams\ sequence by accreting more gas and forming new stars in the outer disks (purple arrow).

A potential objection to this picture is the possible impact that disturbances may have on the structure of galaxies.
The model invokes such events to lower rotational support and trigger central star formation in low-\delSD\ galaxies, but at the same time such events are also credited with building central density and forming bulges \citep{Katz1992, Bournaud2005}. If that is severe enough, such disturbances might increase \delSD\ substantially, leaving no objects behind to populate the so-called ``recovery region'' in the low \delSD\ domain.
Countering this is the fact that many low-\delSD\ galaxies are highly disturbed yet lack high-density centers, and thus mergers may not be universally efficient in making bulges.
Hydro simulations also show that remnants of gas-rich mergers may be disk-dominated with bulge formation suppressed, and disk structures may be produced in later phases of mergers \citep{Robertson2006, Hopkins2009a, Hopkins2009b, Hopkins2010, Moster2010, Keselman2012, Sauvaget2018, Gargiulo2019}. Therefore, low-\delSD\ galaxies may be able to survive these processes without efficient increase in central density, contributing to the distributions seen here.

The mixed scenario contemplated in Fig. \ref{disturbance_diagram} may also not be a complete description of the physical processes.
The key point is that at least two mechanisms seem to be required to understand the correlation of star formation rate and kinematics for low-\delSD\ galaxies, but we do not strictly rule out possibilities of other mechanisms.
One possible candidate is stellar feedback. It has been seen in simulations that strong feedback can drive gas outflow, decrease central star formation and gas dispersion and cease clump formation \citep{El_Badry_2017}. Therefore, galaxies experiencing strong feedback might decrease in $S2$ and increase $\rm D_n4000$ and $\lambda_{R_\mathrm{e}}$, consistent with what is shown in Fig. \ref{struc}.
Stellar feedback would be presented as an arrow with opposite direction to the blue arrow of disturbance in Fig. \ref{disturbance_diagram}, and both processes may play a role in determining the distribution we see in observations.
It would be interesting to further quantify the effects of different mechanisms on galaxy kinematics and star formation state.

\section{Discussion}
\label{Discussion}

\subsection{Evolution on the stellar population vs. central surface density plane}

\begin{figure} 
    \centering
    \includegraphics[width = \columnwidth]{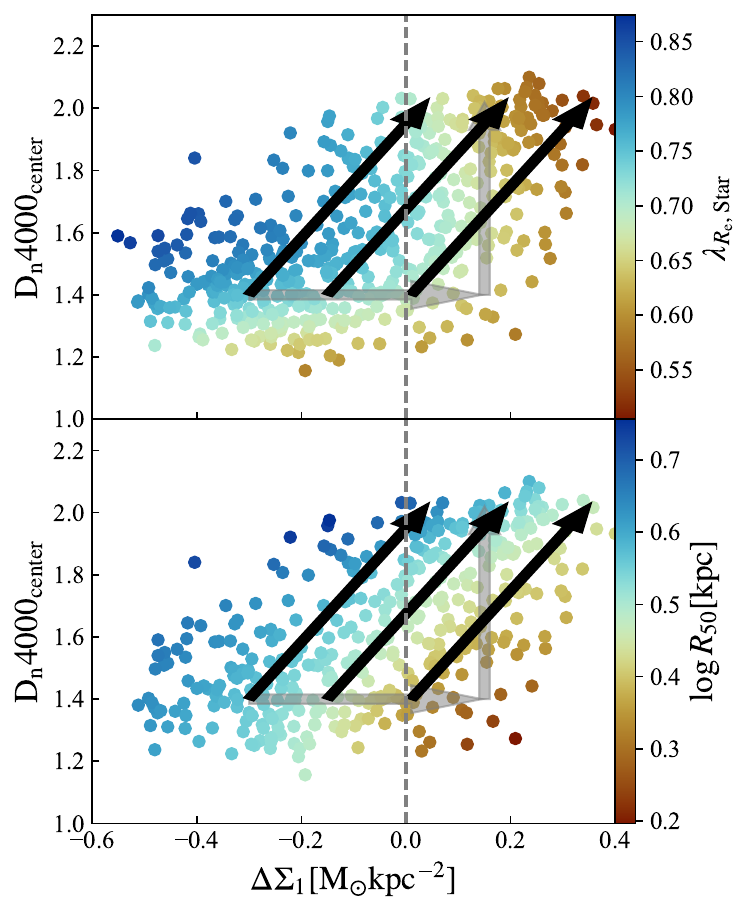}
    \caption{Schematic summary of evolutionary paths on the $\rm D_n4000_{center}$ vs. \delSD\ space. The top panel is colored by LOESS smoothed \lams\ and the bottom panel is colored by smoothed half-mass radii. The gray horizontal and vertical arrows show the hypothesized paths that galaxies first increase relative central densities with stellar population ages almost constant, and then quench with \delSD\ almost unchanged. This scenario is challenged by the effectively reduced $R_{50}$ along the horizontal path, and the black slanted arrows represent an alternative scenario that galaxies maintain their relative ranks of half-mass radii. Under this scenario, the wholesale evolution of pseudo-bulge galaxies into classical-bulge galaxies is not preferred (see text).}
    \label{bulge_evolution}
\end{figure}

We turn now to the question of how galaxies evolve in the parameter space of central $\rm D_n4000$ vs. \delSD.

We get clues to this from the distribution of galaxy effective radius, which is an important second factor that describes galaxy evolutionary state other than stellar mass \citep[e.g.][]{MMW1998,Shen2003,vanderWel2014,Cappellari_2016,Chen_2020}.
It was first interpreted by \citet{vanDokkum_2015} that galaxies may evolve along parallel tracks along $\log R_\mathrm{e}-\rm \log M_*$ when star forming before they meet the boundary of quenching.
The picture was further developed with observations \citep[e.g.][]{Lilly2016,Chen_2020}, and similar evolutionary tracks that galaxies averagely increase $R_\mathrm{e}$ with parallel slopes on $\log R_\mathrm{e}-\rm \log M_*$ are also seen in simulations \citep{Genel_2018} and empirical models \citep{RP2017}.

As the whole distribution of galaxies on $R_\mathrm{e}-\rm M_*$ is also moving to the upper regions with time \citep{vanderWel2014}, a reasonable assumption that combines these properties is that galaxies may maintain their \emph{relative} position in the distribution of radii at fixed mass as they evolve. In other words, galaxies with similar ranks of radii but different masses would be evolutionarily related, and the properties of the massive ones may predict the future of the less massive ones on that rank.

To see how galaxies evolve in central $\rm D_n4000$-\delSD\ under this assumption, we color the points with LOESS smoothed half-mass radii on $\rm D_n4000$-\delSD\ in the bottom panel of Fig. \ref{bulge_evolution}. 
The points in the top panel as complement are colored by LOESS smoothed \lams\ values, same as in the lower right panel of Fig. \ref{Dn4000_center}.
Schematic evolutionary tracks under the assumption of constant radius rankings are represented as the slanted arrows (solid black arrows), along which $R_{50}$ keeps approximately constant. Therefore, if the assumption is true, galaxies would evolve on almost parallel tracks in $\rm D_n4000_{center}$-\delSD\ before they quench, with a strong inclined slope along which both $\rm D_n4000_{center}$ and \delSD\ increase. In contrast, \lams\ evolves non-monotonically, increasing while galaxies are star-forming and then declining when they quench. The early, increasing phases are consistent with the rise in $\lambda_{R_\mathrm{e}}$ with mass (for low-\delSD\ galaxies) in Figs. \ref{lam_Dn4000_m} and \ref{PB_subsets}, and the general picture of rise and then fall vs. mass has been shown in many works \citep[e.g.][]{FB2019, Wang2020,Fraser-McKelvie2021}.

This evolutionary scenario differs from that proposed by \citet{Luo2020}, who argued that galaxies evolve through the elbow shape en route to quenching. 
Their path is shown as the translucent gray arrows in Fig. \ref{bulge_evolution}, along which galaxies increase \delSD\ first with $\rm D_n4000$ almost constant, and then decline in star formation rates with \delSD\ almost unchanged.
\citet{Luo2020} proposed that the first (horizontal) process might result from secular evolution \citep{KK04,Fang_2013}, whereby non-circular mass distributions like bars would torque the gas, causing it to flow to the center, make stars and build $\Sigma_1$ \citep{Wang_2012,Lin_2017,Querejeta2021}. 
However, the new finding in Fig. \ref{bulge_evolution} is that half-mass radii also strongly decrease along that arrow, by almost a full dex (see also Fig. \ref{sample}), and it is not clear whether secular evolution alone could cause such a change.
If secular processes are limited to stars alone, then energy and angular momentum are conserved and some stars move outward as other stars move inwards \citep{Lynden-Bell1972,KK04}, thus causing little net change in half-mass radius.
If gas is brought in along with stars, the ensuing star-formation at the center may cause the half-mass radius to shrink.
However, the overall light distributions of these objects are highly compact; they lack the extended envelope of light that would be produced by conservation of angular momentum during secular evolution.

An alternative process to increase central density is wet mergers, which may be able to cause gas inflow and strong shrinkage in size.
This process, called ``compaction'', is common in simulations of gas-rich galaxies \citep{Dekel2014,Ceverino2015, Zolotov2015, Tacchella2016, Barro_2017} and has been invoked to account for the small radii of quenched galaxies at $z \sim 2$ \citep{Barro_2017}. However, this process is less likely to explain elbow galaxies today. First, gas contents are much lower now than they were at $z \sim 2$, and second, the elbow galaxies are common being the smallest star-forming galaxies at every mass. Compaction may have played a role in their formation in the distant past, but the \citet{Luo2020} arrow in Fig. \ref{bulge_evolution} is meant to indicate today. The notion that large-radii galaxies are evolving en masse to make small galaxies \emph{today} is inconsistent with numerous data, such as the spread in star-forming galaxy radii at fixed mass, which has been remarkably constant since $z \sim 2$ \citep{vanderWel2014}.

We do acknowledge that the black arrows in Fig. \ref{bulge_evolution} are not to be taken literally as accurate evolutionary paths, not only due to the limited sample size but also to the drifting of whole data in all diagrams over time. Even so, the arrows may still be a guide to \emph{how objects will evolve in relation to one another}, and thus to the picture of relative paths in $\rm D_n4000$-\delSD\ space.
It will be interesting to see whether future simulations can reproduce the joint distributions of kinematic properties, star-formation rates, and radii seen today and back in time, and in particular to test the assumption that galaxies with similar radius rankings are evolutionarily linked.

\subsection{The Evolutionary Relationship of Pseudo Bulges and Classical Bulges}
\begin{figure} 
    \centering
    \includegraphics[width = \columnwidth]{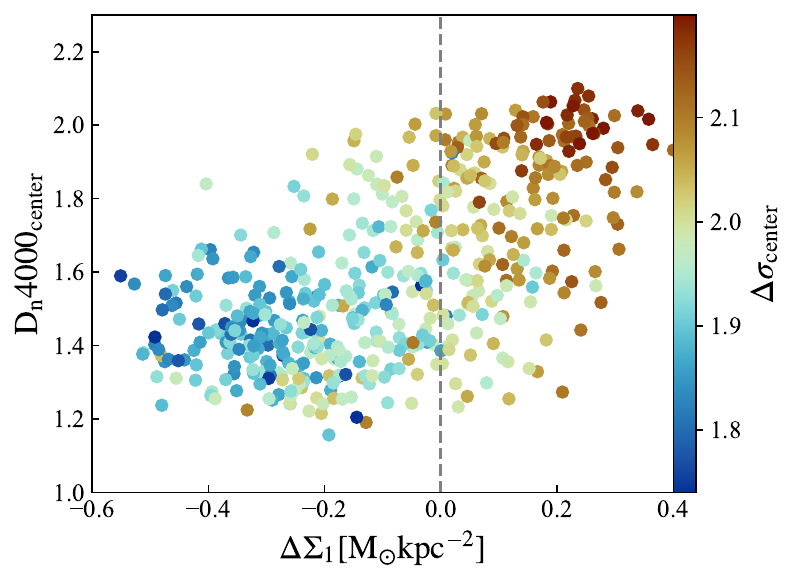}
    \caption{Distribution of a kinematic bulge-indicator, $\rm \Delta \sigma_{center}$ ($\equiv \log \sigma_\mathrm{center} - 0.25(\log \rm M_* - 10.5)$), in $\rm D_n4000_{center}$ vs. \delSD.
    The distribution of $\rm \Delta \sigma_{center}$ is consistent with the J-shape pattern shown by \lams. Most of the red points are to the right of the \delSD\ boundary, with blue points to the left, indicating that \delSD\ does a fairly good job of separating bulges characterized by kinematics. 
    The exception is for the points on the bottom with relatively high $\rm \Delta \sigma_{center}$, located to the left of \delSD\ $= 0$. 
    Nevertheless, if the evolutionary tracks in Fig. \ref{bulge_kinematics} are correct, \delSD\ turns out to be useful in separating bulge types and the wholesale evolution of pseudo-bulge galaxies into classical-bulge galaxies is not preferred (see text).}
    \label{bulge_kinematics}
\end{figure}

We close this discussion on evolution with a brief mention of the possible implications of our findings for the evolution of bulges. This was a major topic in \citet{Luo2020}, which asserted that \delSD\ $= 0$ could be used as the dividing line between pseudo-bulges and classical-bulges and proceeded to argue, as we said above, that pseudo-bulges en masse evolve into the elbow region to become classical bulges. If we take that definition of bulge types, the assumption of constant radii rankings, however, holds an opposite view that most pseudo-bulges will not evolve to classical-bulges in general before they quench.

To this point, we have not mentioned the word ``bulge'' since measurements of bulge type and bulge prominence are rather technical \citep[e.g.][]{KK04,FD2008,Gadotti2009,Kormendy2016}. Moreover, it is probable that the use of \delSD\ $= 0$ as the dividing line is not exact; \citet{Luo2020} compared \delSD\ with more precise bulge measurements by \citet{Gadotti2009}, and while the agreement was good, it was not perfect. But MaNGA data add dynamical estimates of bulge prominence to the space of $\rm D_n4000$ vs. \delSD, permitting us to take a new look at the accuracy of \delSD\ in classifying bulges and at the same time reconsider whether pseudo bulges evolve to classical bulges.

Fig. \ref{bulge_kinematics} plots a simple dynamical parameter, $\log \sigma_{\rm center}-0.25\log \rm M_*$ in $\rm D_n4000$ vs. \delSD\ to indicate bulge types, as central dispersion should be sensitive to bulge prominence \citep{KK04,Fabricius_2012, Zhao2012, FD2016, Sachdeva2020, Hu2024}.
This parameter removes the general trend of all velocities to increase roughly as $\rm M_*^{0.25}$. As shown in Fig. \ref{bulge_kinematics}, classical-bulge candidates (high \delSD) have relative high central dispersions, while pseudo-bulge candidates (low \delSD) have relatively low central dispersions.

This new color pattern strongly resembles the J-shaped pattern in Fig. \ref{Dn4000_center}, which is not unexpected since $\lambda_{R_\mathrm{e}}$ and velocity dispersion should correlate inversely. Furthermore, the dividing line \delSD\ $= 0$ actually does a reasonable job to first order of separating bulge candidates indicated by kinematics: classical-bulges (red) lie generally to the right and pseudo-bulges (blue) lie generally to the left. The boundary of \delSD\ $= 0$ corresponds to $\Delta \rm \sigma_{center} \sim 2$, consistent with the classifications in \citet{Sachdeva2020} and \citet{Hu2024}. In more detail, however, a population of relatively high-$\Delta \sigma$ galaxies is seen to run along the lower edge of the distribution on the left-hand side.
These points are on the ``wrong'' side of \delSD. They lie near the bottom of the middle arrow in Fig. \ref{bulge_evolution} and consist of low-mass star-forming galaxies with intermediate radii and borderline high relative central $\sigma$. If the arrows in Fig. \ref{bulge_evolution} are correct, their bulge prominence (characterized by kinematics) will actually \emph{decline} in future as they gain new mass in the form of dynamically cold disks (i.e., they are low-mass galaxies on the ridgeline in Fig. \ref{lam_Dn4000_m}). These objects do not fit the profile of ``classical bulges'' as commonly understood, and if they are ignored, then \delSD\ $= 0$ turns out to be a useful dividing line between the two types of bulges.
Regardless, however, the main point still stands: if the tilted arrows in Fig. \ref{bulge_evolution} are correct, some borderline pseudo-bulges may become classical bulges, but the wholesale evolution of all pseudo-bulges to classical bulges does not appear to be correct.

\subsection{Caveats}
\label{Caveats}
\subsubsection{Lack of spectral resolution for low dispersion}
Learning kinematics of disk galaxies is limited by spectral resolution. The instrumental dispersion of MaNGA survey is $\sim 67 \mathrm{km/s}$, and the stellar dispersion measurement is suggested to be reliable down to 50km/s when S/N > 10 \citep{Westfall_2019}. However, many studies have found that dispersion may be overestimated when lower than instrumental dispersion \citep[e.g.][]{FB2017}. Dispersion measurement is biased and the measured $\lambda_{R_\mathrm{e}}$ is a ``lower limit'' of the true values. This could introduce artificial effects on the relations we see.

A check on our results is whether they are similar in both gas and stellar kinematics.
The instrumental effect is less severe for gas measurements, since the S/N of emission line can be much higher than that of continuum for star forming galaxies. 
A comparison between MaNGA and DiskMass survey ($R \sim 10000$) \citep{Bershady2010} shows that for MPL10 and later versions the observed $\rm H\alpha$ dispersion is not significantly biased in the region of $4 \arcsec  < r < 15 \arcsec$, where typical H$\alpha$ dispersion are $\sim 20 \rm km/s$ and are much lower than the instrumental dispersion \citep{Law_2021}. Therefore, if the results of stellar kinematics are reproduced in gas kinematics, they are likely to be real.

We checked the results for gas kinematics and show the figures in Appendix \ref{gas_kinematics}. 
The patterns of $\lambda_{R_\mathrm{e}}$ on $\rm D_n4000$-\delSD\ plot are both seen for stellar and gas kinematics and we propose they are likely to be correct. Though the relation between mass and $\lambda_{R_\mathrm{e}}$ for low-\delSD\ galaxies is not clear for gas (Fig. \ref{lambda_m_Dn4000_g}), it could be true given that gas is dissipational and more settled. In previous studies of SAMI and CALIFA, similar dependence of stellar $\lambda_{R_\mathrm{e}}$ on mass was seen \citep{FB2019, Fraser-McKelvie2021, vandeSande2021a, vandeSande2021b, Cortese2022}, and we believe it is real.
In conclusion, all of our important conclusions ARE reproduced in stars and gas if appropriate allowance is made for the greater dissipational behavior of gas.
While these results should be further validated through future tests with higher resolution, we have confidence in their overall validity.

\subsubsection{Lack of spatial resolution for small galaxies}
Another important effect in kinematic measurements is from the limited spatial resolution. Low spatial resolution can flatten velocity and dispersion profiles, resulting in underestimation of $\lambda_{R_{\rm e}}$. We limit the sample to have $R_{\rm e} > 5\arcsec$ and do not apply any correction.
In Appendix \ref{spatial_resolution} we show that under our sample selection, the effect of low spatial resolution is not severe and could be neglected without correction.

\subsubsection{Selection bias for high-\delSD\ galaxies}
Although the restriction of the sample to objects with $R_\mathrm{e} > 5\arcsec$ mitigates the spatial resolution problem, the selection itself can make the sample biased.
The excluded small galaxies ($R_{\rm e} < 5\arcsec$) only make up a relatively small fraction ($\sim 25\%$) of low-\delSD\ galaxies, but a considerable fraction ($\sim 67\%$) of high-\delSD\ galaxies.
The selection bias does not largely affect our conclusions of low-\delSD\ galaxies, but makes it hard to interpret evolution for high-\delSD\ galaxies. Moreover, as indicated in Fig. \ref{sample}, high-\delSD\ galaxies is a composite category with a large range of star formation rates. We do not give detailed interpretations of high-\delSD\ galaxies in this paper, and plan to address these issues in a future work (Wang et al. in prep). 

\section{Conclusions}
\label{Conclusions}
In this paper we show how galaxy rotational support varies as a function of stellar population age and central surface densities with $\sim 500$ galaxies from MaNGA.
With \delSD\ and $\mathrm{D_n4000}$ as structural and stellar population indicators, we show that the rotational support correlates with central and global properties in a complex manner.
The main results are as follows:
\begin{enumerate}
    \item We confirm the conclusions of many previous works that rotational support (\lams) statistically decreases with relative central density (\delSD) and stellar population age ($\rm D_n4000$) with some mass dependence as well (Figs. \ref{Dn4000_center}, \ref{Dn4000_global}). For the first order, \delSD\ and $\rm D_n4000$ increase with stellar mass with \lams\ decreasing.
   
    \item The above trends exist in a large smooth landscape, in which rotational support increases with central/global stellar population age when \delSD\ is low and decreases with \delSD\ and global stellar population when \delSD\ is high (Fig. \ref{Dn4000_center}).
    This landscape is visible as a smooth ``J-shape'' pattern on the 2-D landscape of central properties, $\rm D_n4000_{center}$ vs. \delSD, and as smooth tilted stripes (a fainted ``J-shape'') on the landscape of global $\rm D_n4000$ vs. \delSD.
    Specifically, in low-\delSD\ galaxies, $\lambda_{R_\mathrm{e}}$ increases with both central and global $\rm D_n4000$, but shows a stronger correlation with central stellar population properties (Figs. \ref{Dn4000_center}, \ref{Dn4000_global}).
    For high-\delSD\ galaxies, $\lambda_{R_\mathrm{e}}$ decreases more with global $\rm D_n4000$ but shows a weak dependence on central $\rm D_n4000$ (Figs. \ref{Dn4000_center}, \ref{Dn4000_global}). These specific features are largely independent of stellar mass.
    The new patterns indicate that a combination of stellar populations and structures is needed to characterize the rotational support more completely. 

    \item For low-\delSD\ galaxies, plots of \lams\ vs. mass and central $\rm D_n4000$ show a ridgeline of morphologically normal galaxies plus points scattered to lower $\lambda_{R_\mathrm{e}}$ by disturbances (Figs. \ref{Dn4000_center}-\ref{PB_subsets}).
    Along the ridgeline $\lambda_{R_\mathrm{e}}$ increases with central stellar population ages and stellar mass.
    The low-\lams\ outliers are either morphologically or kinematically disturbed, and the interacting galaxies with strong mergers in process show markedly lower \lams\ (Figs. \ref{PB_subsets}, \ref{struc}).
    Moreover, galaxies with disturbed morphologies or strong mergers also tend to have low central $\rm D_n4000$ (Figs. \ref{PB_subsets}, \ref{struc}, \ref{PB_gallery}).

    \item For low-\delSD\ galaxies, the properties described in (iv) and their relationships are consistent with a ``mixed'' evolutionary model (Fig. \ref{disturbance_diagram}).
    In this model, disturbance effects are superimposed on an underlying evolutionary trend where \lams\ and $\rm D_n4000_{center}$ increase with mass.
    The increasing trend of rotational support with mass, where galaxies get kinematically and morphologically settled, is consistent with a picture in which galaxies become more and more settled as they grow in mass.
    At a fixed mass, galaxies with low $\lambda_{R_\mathrm{e}}$ may have experienced mergers or instabilities, which introduce disturbance, feed central star formation with accreted gas and flatten $\rm D_n4000$ profiles.

    \item Low-\delSD\ galaxies have systematically larger half-mass radii than high-\delSD\ galaxies at a fixed mass (Fig. \ref{sample}). Under the assumption that galaxies have maintained their relative rankings in radius in recent times, we plot schematic evolutionary paths for star-forming galaxies in the space of $\rm D_n4000_{center}$ vs. \delSD\ (Fig. \ref{bulge_evolution}). Galaxies with different radii follow steep paths in which $\rm D_n4000_{center}$ increases greatly as they age whereas \delSD\ increases only modestly. Applying this conclusion to different bulge types, we infer that pseudo-bulges and classical bugles are on different evolutionary paths and remain separate at least until quenching (Figs. \ref{bulge_evolution}, \ref{bulge_kinematics}).
\end{enumerate}

The results in this paper add more information substantially to this existing body of data, which already show the complex behaviours and interplay of structural, stellar population and kinematic properties.
In the future, simulations and observations with high resolution can be useful for understanding dynamical processes and star formation histories in galaxy evolution. Upcoming IFS surveys in the future can also provide more detailed information about galaxy spatial stellar population and kinematics. It would be important to investigate the physical processes that produce the observed distributions, and to test the assumption where galaxies maintain their relative ranks of radii in future simulations.

\section*{Acknowledgements}
This work is partly supported by the National Key Research and Development Program of China (No. 2018YFA0404501 to SM), by the National Natural Science Foundation of China (Grant No. 11821303 to SM).
We would like to thank Drs. Dandan Xu and Song Huang and an anonymous referee for constructive and insightful suggestions and comments which greatly improved the paper. 
We express appreciation for the scientific colormaps provided by cmcrameri and cmasher, and acknowledge the use of the TOPCAT\footnote{http://www.starlink.ac.uk/topcat/} plotting and data analysis tool, which proved indispensable.  

Funding for the Sloan Digital Sky Survey IV has been provided by the Alfred P. Sloan Foundation, the U.S. Department of Energy Office of Science, and the Participating Institutions. SDSS-IV acknowledges support and resources from the Center for High Performance Computing  at the University of Utah. The SDSS website is www.sdss4.org. SDSS-IV is managed by the Astrophysical Research Consortium for the Participating Institutions of the SDSS Collaboration including the Brazilian Participation Group, the Carnegie Institution for Science, Carnegie Mellon University, Center for Astrophysics | Harvard \& Smithsonian, the Chilean Participation Group, the French Participation Group, Instituto de Astrof\'isica de Canarias, The Johns Hopkins University, Kavli Institute for the Physics and Mathematics of the Universe (IPMU) / University of Tokyo, the Korean Participation Group, Lawrence Berkeley National Laboratory, Leibniz Institut f\"ur Astrophysik Potsdam (AIP),  Max-Planck-Institut f\"ur Astronomie (MPIA Heidelberg), Max-Planck-Institut f\"ur Astrophysik (MPA Garching),Max-Planck-Institut f\"ur Extraterrestrische Physik (MPE), National Astronomical Observatories of China, New Mexico State University, New York University, University of Notre Dame, Observat\'ario Nacional / MCTI, The Ohio State University, Pennsylvania State University, Shanghai Astronomical Observatory, United Kingdom Participation Group, Universidad Nacional Aut\'onoma de M\'exico, University of Arizona, University of Colorado Boulder, University of Oxford, University of Portsmouth, University of Utah, University of Virginia, University of Washington, University of Wisconsin, Vanderbilt University, and Yale University.

J.G.F-T gratefully acknowledges the grant support provided by Proyecto Fondecyt Iniciaci\'on No. 11220340, and from the Joint Committee ESO-Government of Chile 2021 (ORP 023/2021), and from Becas Santander Movilidad Internacional Profesores 2022, Banco Santander Chile. 

NFB acknowledges Science and Technologies Facilities Council (STFC) grant ST/V000861/1.

\section*{Data availability}
The data presented in this paper are available upon request for legitimate scientific purposes.


\bibliographystyle{mnras}
\bibliography{MaNGAbulges} 


\appendix
\label{Appendix}

\section{mass-weighted vs. luminosity-weighted rotational supports}
\label{lambda_mv}
\begin{figure}
    \centering
    \includegraphics[width = \columnwidth]{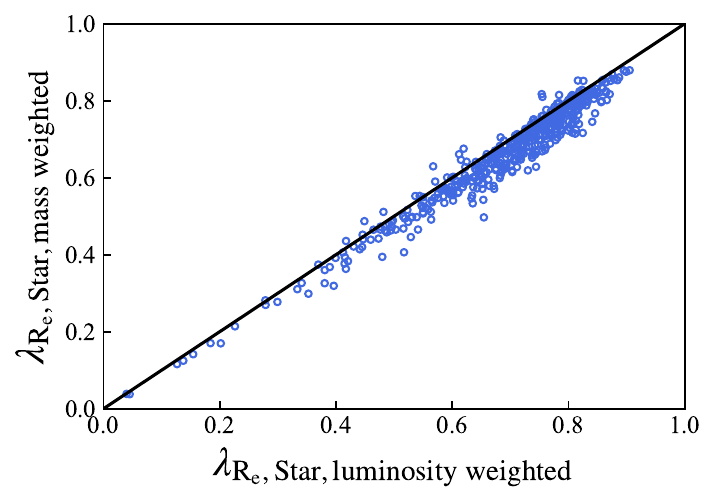}
    \caption{Mass-weighted \lams\ vs. light-weighted \lams. The luminosity weighted \lams\ is systematically higher. A reasonable interpretation is that the outer disk component tends to have more star formation and lower mass-to-light ratio. The difference between two \lams\ values generally increases as \lams\ increases. The difference is up to $\sim 0.1$ and for $\sim 85\%$ galaxies the value is up to $\sim 0.06$. The difference does not affect our analysis.}
    \label{lam_mw}
\end{figure}

We show a comparison between mass-weighted $\lambda_{R_\mathrm{e}}$ and light-weighted $\lambda_{R_\mathrm{e}}$ in Figure \ref{lam_mw}, to investigate the potential biases introduced by factors like disk fading. The mass maps are from the MaNGA FIREFLY value-added catalog \citep{Neumann2022}.
In Fig. \ref{lam_mw}, for $\sim 85\%$ galaxies, the difference between these two values is less than $\sim 0.06$. We also performed tests employing mass-weighted $\lambda_{R_\mathrm{e}}$ and found that our conclusions remain unaffected.

\section{The effect of beam smearing on kinematic measurements}
\label{spatial_resolution}

\begin{figure} 
    \centering
    \includegraphics[width = \columnwidth]{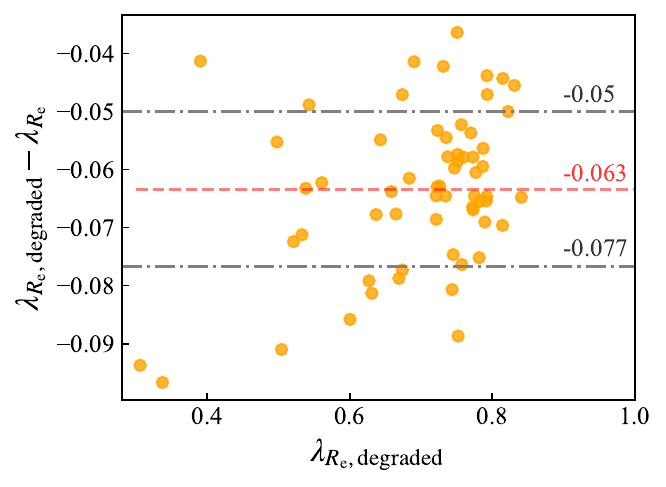}
    \caption{The difference between degraded and original \lams\ vs. degraded \lams\ for the 61 galaxies with $R_\mathrm{e} > 10\arcsec$ in our sample. The red and gray dashed lines indicate the mean value and 1$\sigma$ scatter of the y-axis. The difference between degraded and original values has a mean value of $\sim -0.063$ with 1 $\sigma$ of $\sim 0.013$, and shows no dependence with the degraded values.}
    \label{lam_seeing}
\end{figure}

In this section, we show that under our specific sample selection criteria (galaxies with $R_\mathrm{e} > 5\arcsec$), the effect of beam smearing is not severe and does not bias the main conclusions.
To estimate the beam smearing effect on $\lambda_{R_{\rm e}}$, we adopt a simulation method which is similar to the method in \citet{Greene2018}. We begin by selecting 61 galaxies with $R_\mathrm{e}$ larger than $10 \arcsec$ from our primary sample.
Given that the full width at half maximum (FWHM) of the Point Spread Function (PSF) for MaNGA is $\sim 2.5\arcsec$, which is half of the lower limit of $R_\mathrm{e}$ in our sample, we degrade the spatial resolution of these 61 galaxies to $R_\mathrm{e}$/2, by convolving with a Gaussian PSF with FWHM = $\sqrt{(0.5R_\mathrm{e})^2 - \rm FWHM_{int}^2}$, where $R_\mathrm{e}$ and FWHM are both in units of arcsec. The spectrum of each pixel is simulated as a continuum plus a single Gaussian-shaped absorption line by adopting $V$ and $\sigma$ in MaNGA DAP, and the degraded $V$ and $\sigma$ maps are generated by Gaussian-fitting to the new spectra.

Fig. \ref{lam_seeing} shows the difference of the degraded and the original $\lambda_{R_\mathrm{e}}$ values.
Under the sample selection of $R_\mathrm{e} > 5\arcsec$, the effect of beam smearing in underestimating \lams\ is $\sim 0.063$ with a maximum at $\sim 0.09$, and has no dependence on the degraded values. Therefore, the degradation on \lams\ caused by spatial resolution is small with no bias on degraded values, and does not affect our main conclusions.

\section{Weighting of spectral indices and emission line equivalent widths}
\label{weighting}

\begin{figure}
    \centering
    \includegraphics[width=0.5\textwidth]{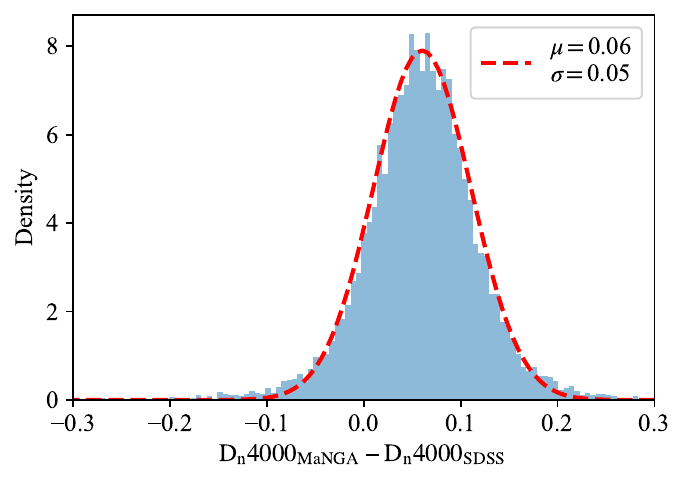}
    \caption{Histogram of the difference for $\rm D_n4000$ between MaNGA central measurements within radii of 1.5$\arcsec$ and SDSS fiber ($r=1.5\arcsec$) measurements. The sample is the original MaNGA sample crossed matched with MPA-JHU measurements, $\sim 9000$ galaxies. The y axis is density, and the labels are the mean and standard deviation of the best-fit Gaussian. The mean value of the difference is $\sim 0.06$, comparable to $\sigma$. The offset in $\rm D_n4000$ is also seen in \citet{Westfall_2019}, and we infer that it is caused by emission line subtraction processes for each pixel.}
    \label{MaNGA_SDSS}
\end{figure}

The spectral indices measured within apertures are luminosity-weighted mean values, where the pixel weightings are provided in MPL11 as \texttt{SPECINDEX\_WGT}. To assess the validity of measuring these mean values directly from maps instead of stacking individual spectra, a comprehensive evaluation is presented in \citet{Westfall_2019}. In this study, we make a comparison of our $\rm D_n4000$ measurements for MaNGA galaxies with those derived from integrated spectra by MPA-JHU, as illustrated in Figure \ref{MaNGA_SDSS}.
An offset of $\sim 0.06$ is observed, which is on the order of the standard deviation ($\sigma$) in Gaussian fitting procedures. It is noteworthy that this offset is also documented in \citet{Westfall_2019}. We infer that it is potentially attributed to the emission line subtraction process for each pixel. The observed difference is relatively minor and does not introduce significant impact on our results.

\section{Rotational support for $\rm H\alpha$ emission lines}
\label{gas_kinematics}
\begin{figure*}
    \centering
    \includegraphics[width=\textwidth]{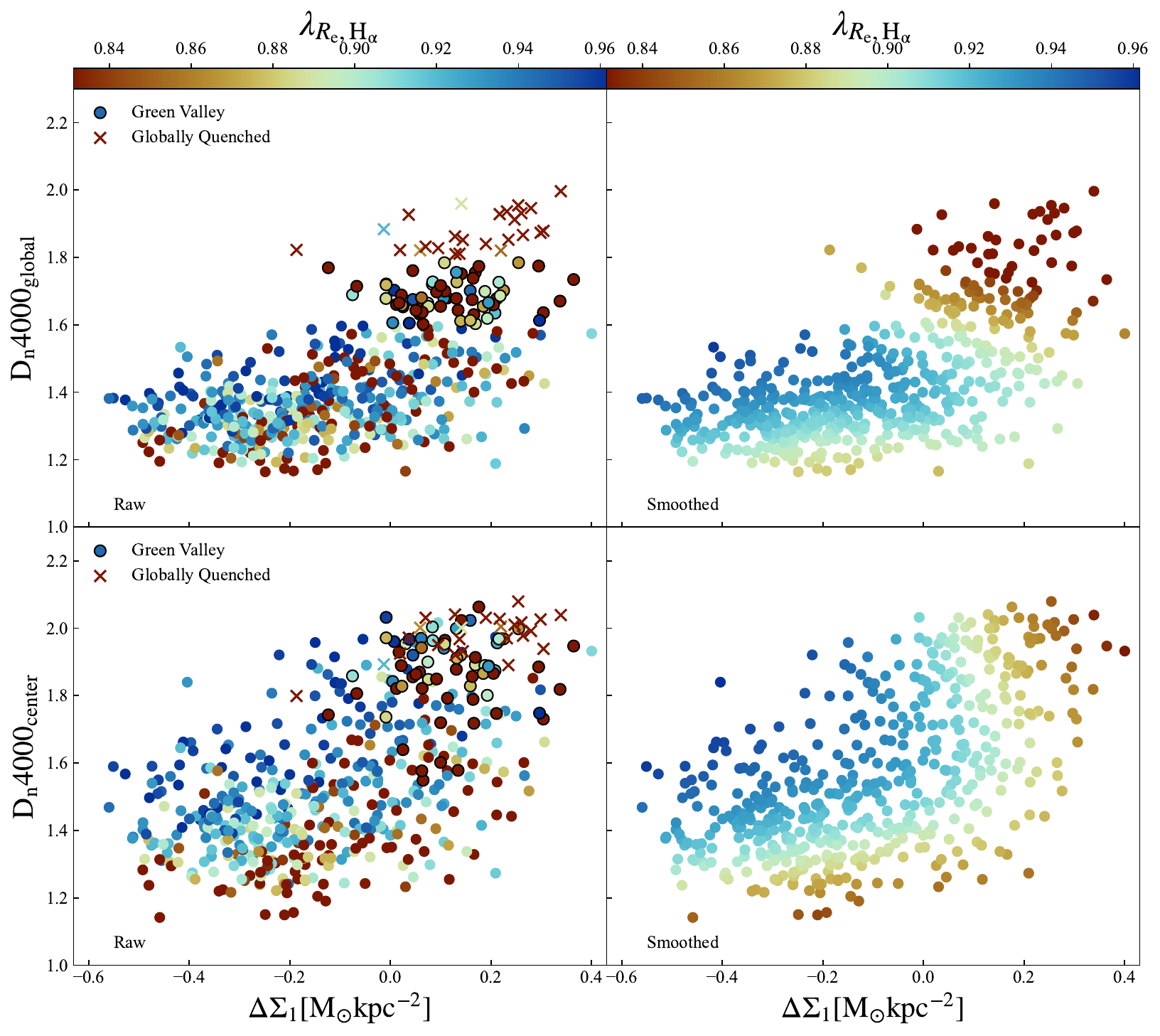}
    \caption{$\mathrm{D_n4000_\mathrm{global}}$ and $\mathrm{D_n4000_{center}}$ vs. \delSD\ colored by \lamg. The left column shows the original scatter plots and the right panels are LOESS smoothed plots. Fully quenched galaxies ($\mathrm{D_n4000_\mathrm{global}} > 1.8$) are labeled as crosses and galaxies in green valley ($1.6 < \mathrm{D_n4000_\mathrm{global}} < 1.8$) as circles with black edges.
    The sample is based on gas data quality (SNR of $\rm H\alpha$ emission) and is different from Fig. \ref{Dn4000_center}.
    The distribution of \lamg\ shows striped pattern in global $\rm D_n4000$ vs. \delSD\ and ``J-shaped'' pattern in central $\rm D_n4000$ vs. \delSD, similar to what is shown in Fig. \ref{Dn4000_center}.}
    \label{Dn4000_center_g}
\end{figure*}

\begin{figure*}
    \centering
    \begin{subfigure}{\columnwidth}
        \includegraphics[width = \columnwidth]{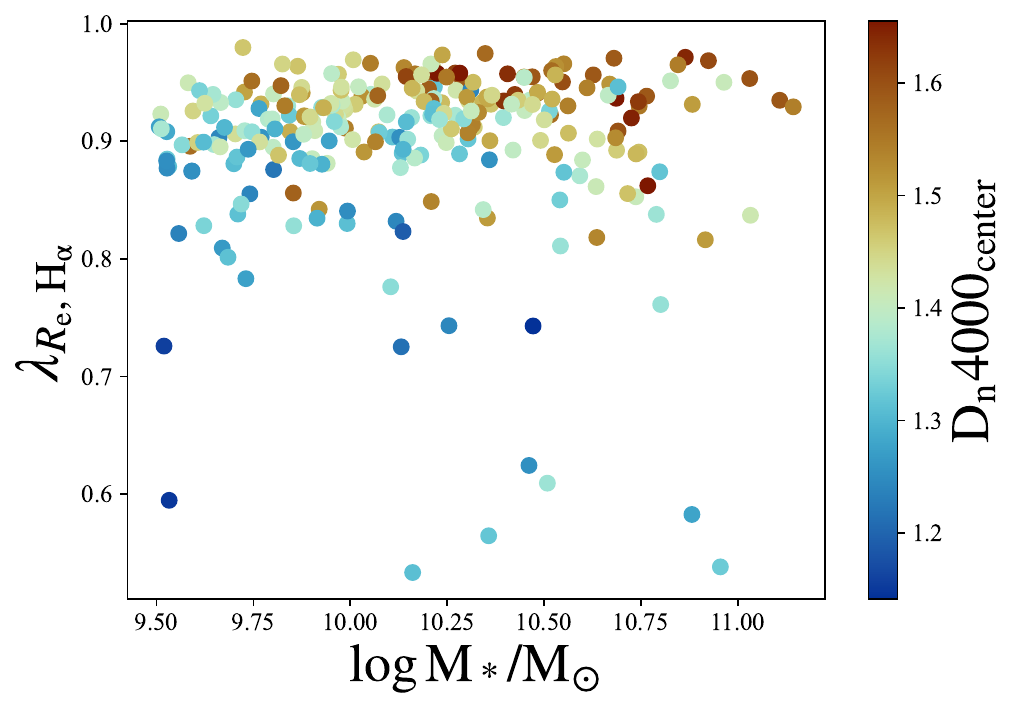}
        \caption{\lamg\ vs. stellar mass colored by central $\rm D_n4000$.}
        \label{lambda_m_Dn4000_g}
    \end{subfigure}
    \begin{subfigure}{\columnwidth}
        \includegraphics[width = \columnwidth]{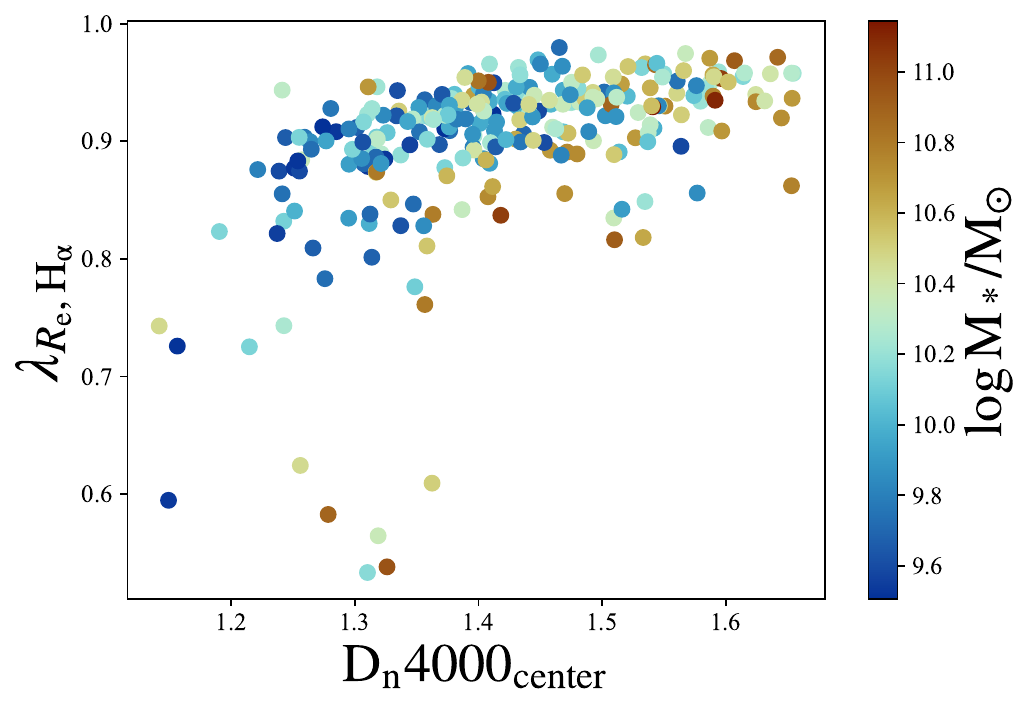}
        \caption{\lamg\ vs. central $\rm D_n4000$ colored by stellar mass.}
        \label{lambda_Dn4000_m_g}
    \end{subfigure}
    \caption{Replots of Fig. \ref{lam_Dn4000_m} with \lamg. Only galaxies with negative \delSD\ and $\rm D_n4000_{center} < 1.66$ are shown. The sample is based on gas data quality (SNR of $\rm H\alpha$ emission) and is different from Fig. \ref{lam_Dn4000_m}.}
\end{figure*}

We test our main results with \lamg, which is measured for $\rm H\alpha$ emission lines. Figs. \ref{Dn4000_center_g}, \ref{lambda_m_Dn4000_g} and  \ref{lambda_Dn4000_m_g} are identical to Figs \ref{Dn4000_center}, \ref{lambda_m_Dn4000} and \ref{lambda_Dn4000_m} with \lamg\ measurements in substitution. \lamg\ is systematically higher than \lams\ owing to the well known trend of gas, which is dissipative, to be dynamically ``colder'' than stars. Aside from this, the general trends consistent with stellar kinematics are reproduced.

\section{Sample maps of galaxies from different locations of stellar population-structure parameter space}
\label{maps}

\begin{figure*}
    \centering
    \includegraphics[width=\textwidth]{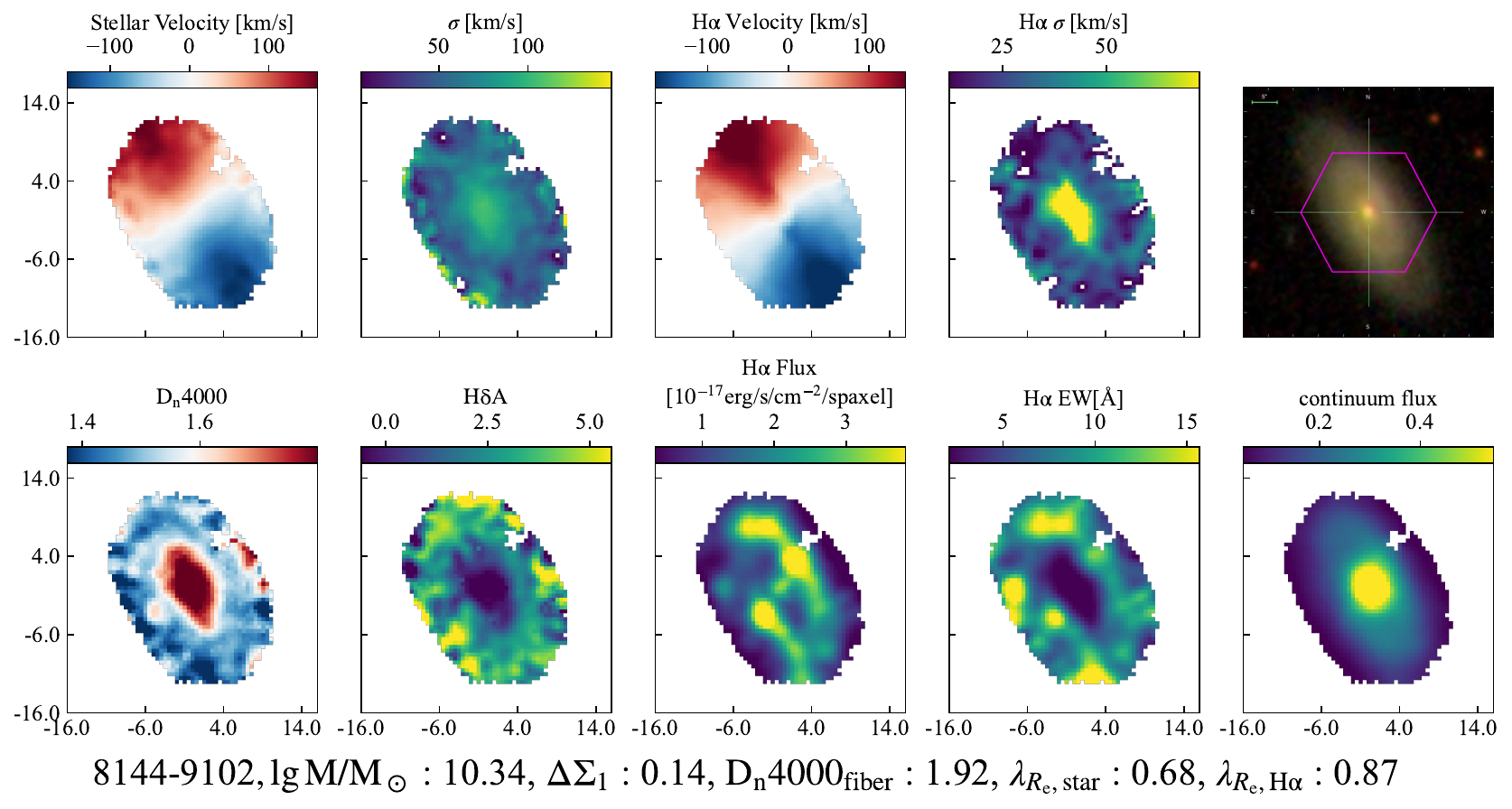}
    \caption{8144-9102, an example of centrally quiescent galaxies with high \delSD. This galaxy is also flagged as having ``perturbed kinematics''.}
    \label{QC}
\end{figure*}
\begin{figure*}
    \centering
    \includegraphics[width=\textwidth]{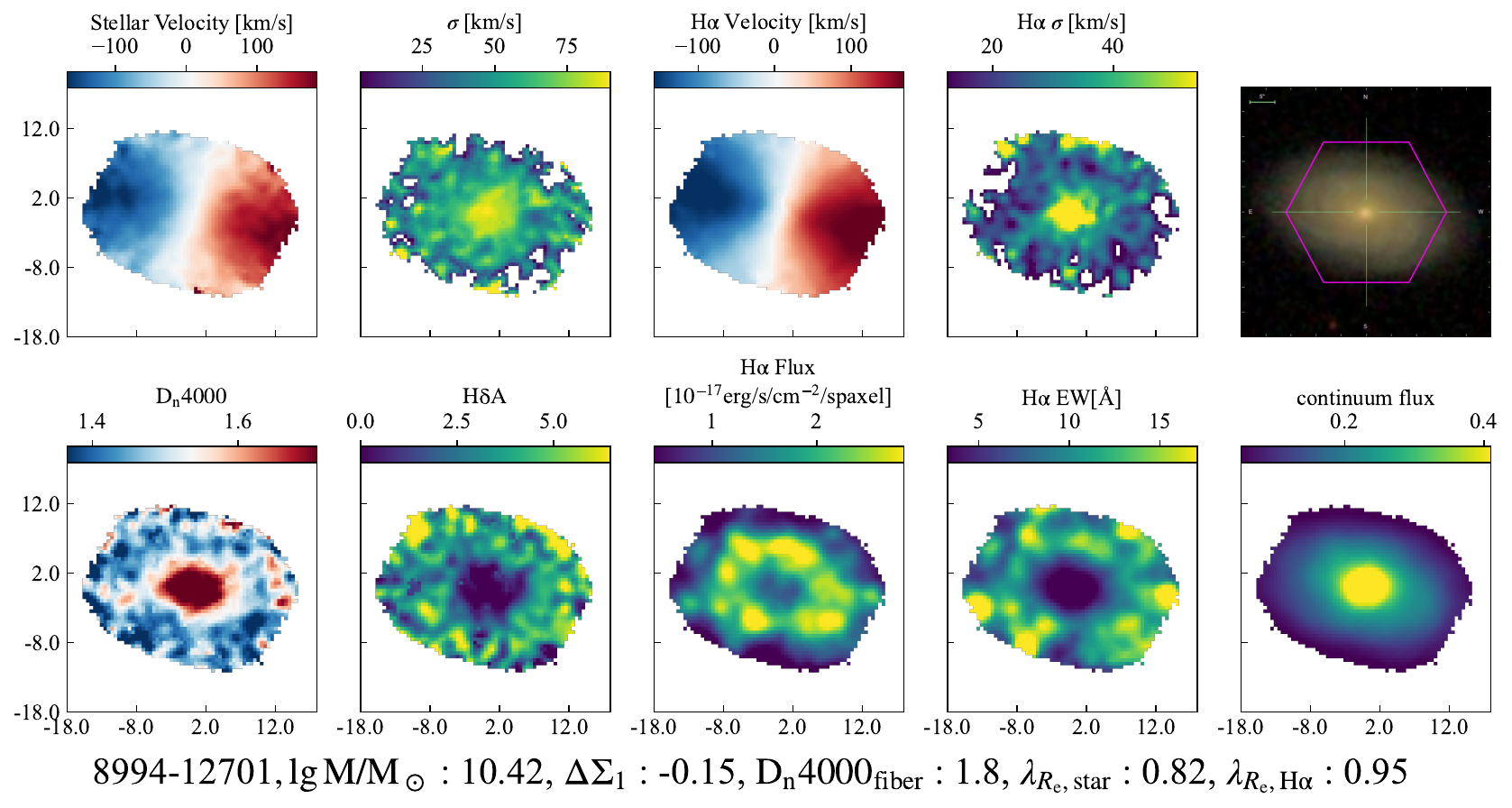}
    \caption{8994-12701, an example of centrally quiescent galaxies with low \delSD.}
    \label{QP}
\end{figure*}
\begin{figure*}
        \centering
        \includegraphics[width=\textwidth]{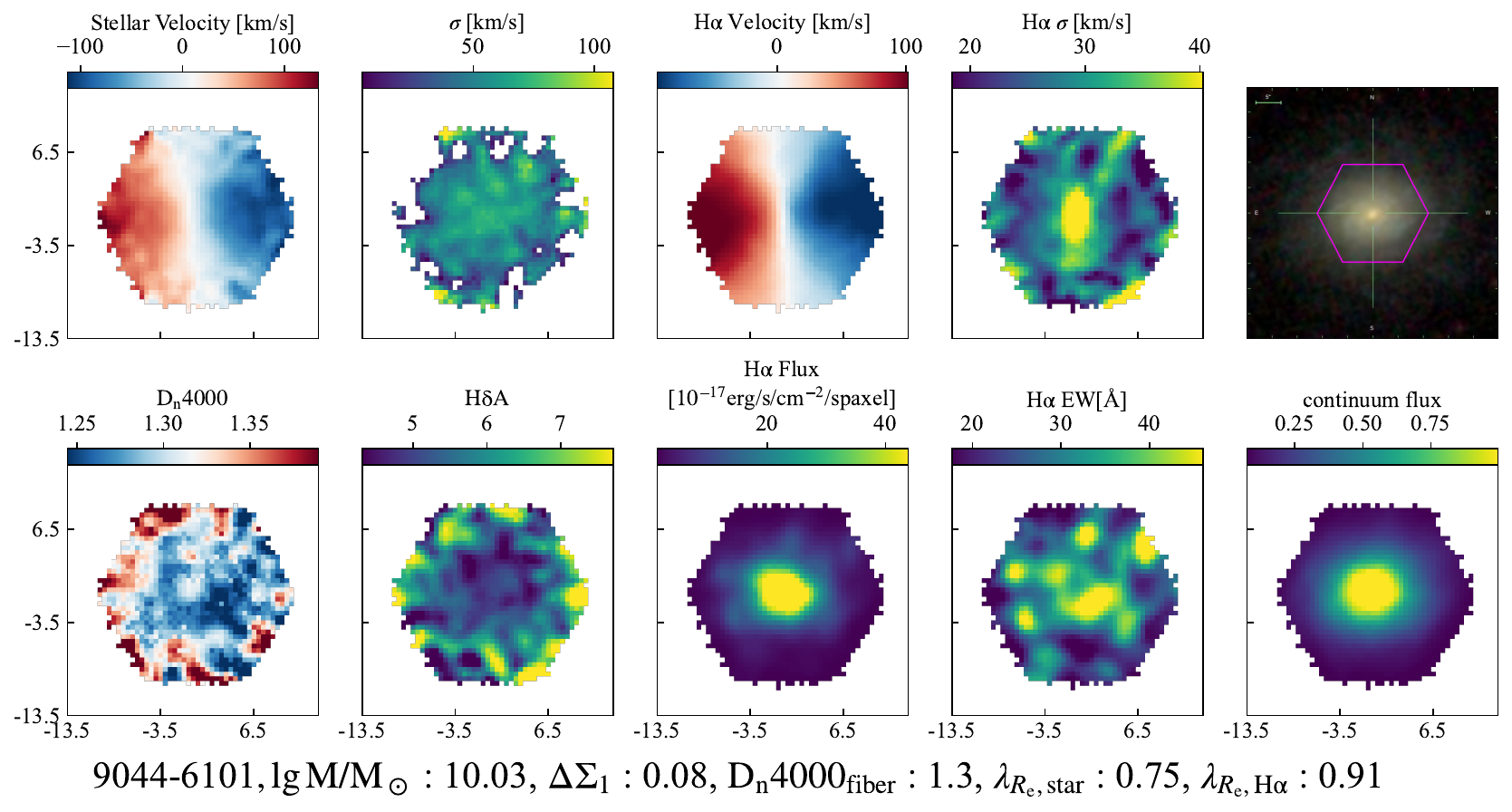}
        \caption{9044-6101, an example of centrally star-forming galaxies with high \delSD.}
    \label{SFC}
\end{figure*}
\begin{figure*}
    \centering
    \includegraphics[width=\textwidth]{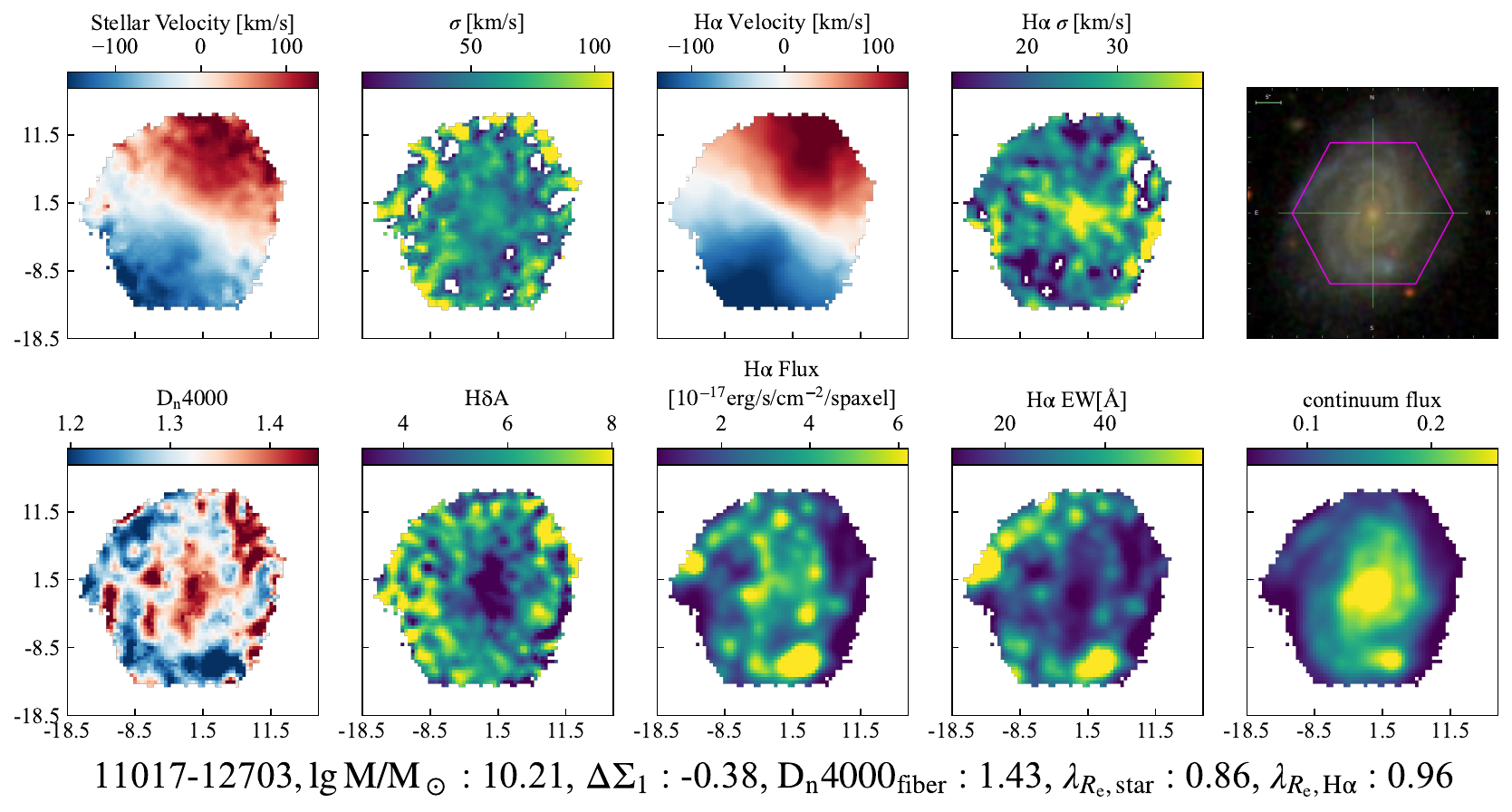}
    \caption{11017-12703, an example of centrally star forming galaxies with low \delSD.}
    \label{SFP}
\end{figure*}
\begin{figure*}
    \centering
    \includegraphics[width=\textwidth]{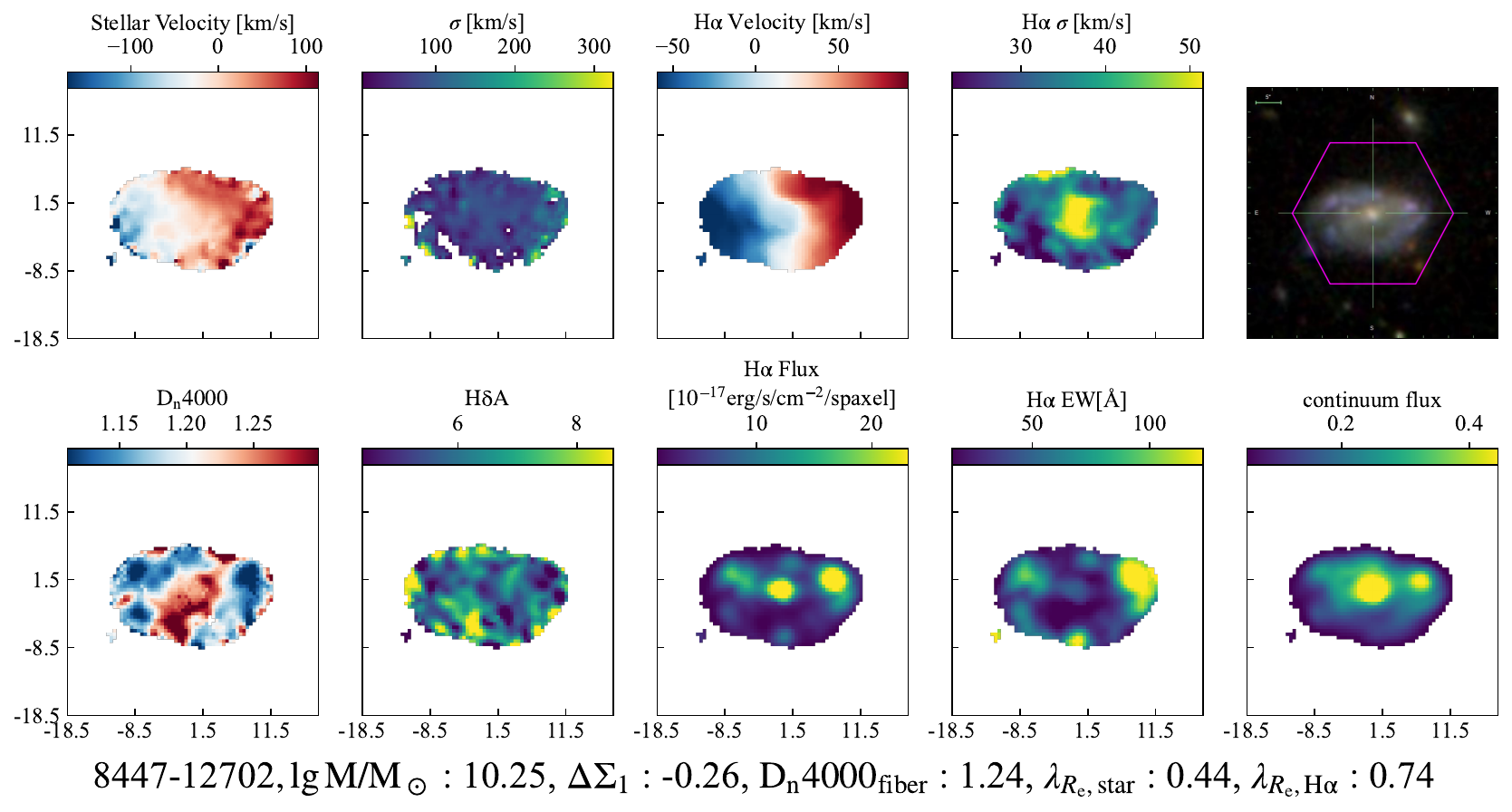}
    \caption{8447-12702, an example of galaxies with low \delSD\ and very young stellar population ($\rm D_n4000_{center} < 1.3$). This galaxy is also flagged as having ``perturbed kinematics''.}
    \label{SRP}
\end{figure*}

In this section, we present maps of kinematics, flux, and spectral indices for sample galaxies in Figures \ref{QC} through \ref{SRP}. Galaxies are classified into 4 types based on their locations on $\mathrm{D_n4000_{center}}-\Delta \Sigma_1$ plot: centrally star-forming galaxies with high \delSD\ (lower right), centrally star forming galaxies with low \delSD\ (lower left), centrally quiescent galaxies with high \delSD\ (higher right) and centrally quiescent galaxies with low \delSD\ (higher left).  We also show sample maps of low-\delSD\ galaxies with $\mathrm{D_n4000_{center}} < 1.3$, as they tend to have less rotational support and more disturbed morphologies.

\section{Gallery of low-central-surface-density galaxies with different central stellar population ages}

\begin{figure*} 
    \centering
    \includegraphics[width = \textwidth]{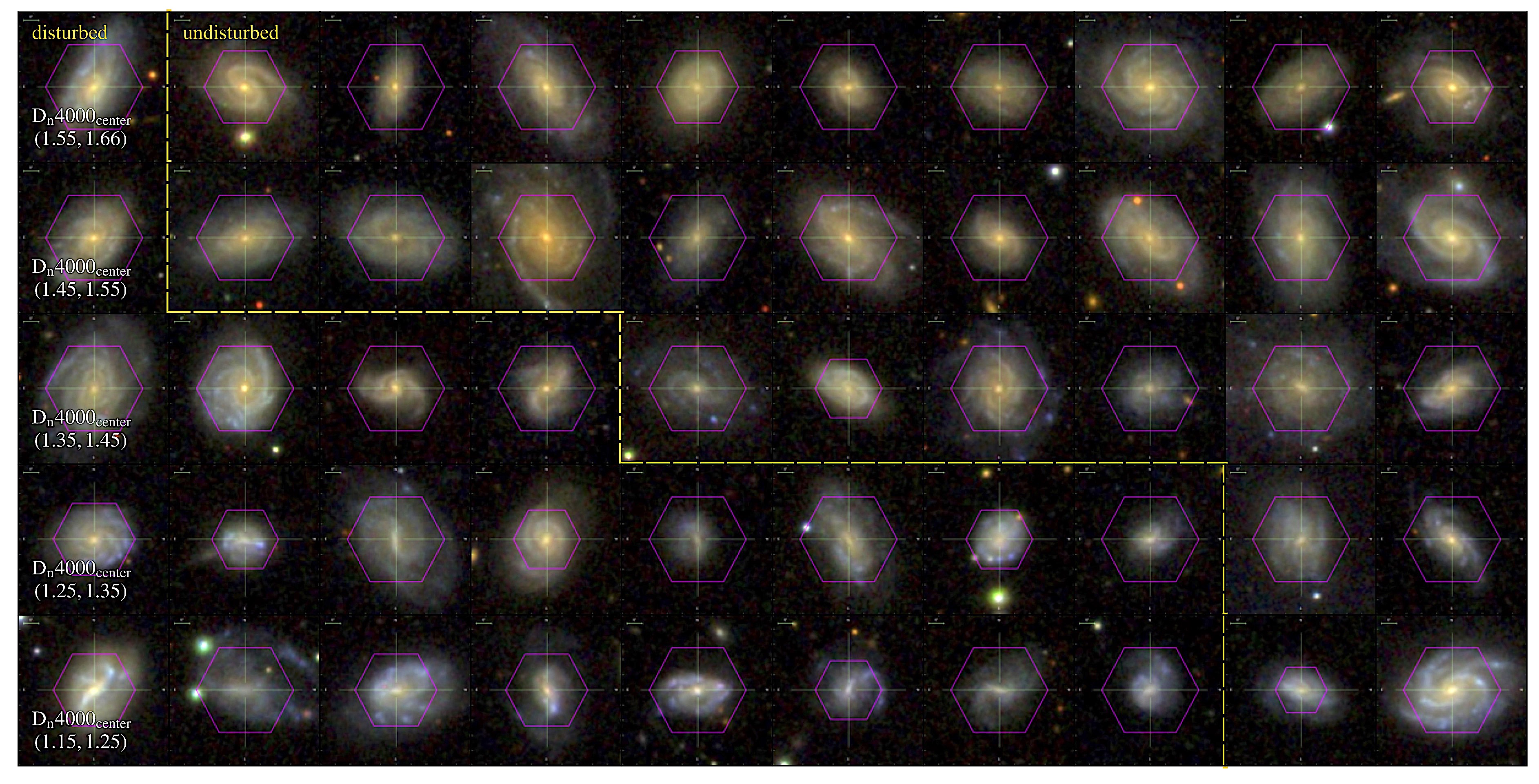}
    \caption{An image gallery of low-central-surface-density galaxies (\delSD\ < 0) with different central stellar populations and degrees of disturbance. From bottom to top are images of galaxies with $\mathrm{D_n4000_{center}}$ of five bins, (1.15, 1.25), (1.25, 1.35), (1.35, 1.45), (1.45, 1.55), (1.55, 1.66). The images left to the orange dashed line are galaxies with disturbed morphology (Sec. \ref{sample_selection}, black triangles in Fig. \ref {PB_subsets}), and the images on the top-right are from the clean sample (Sec. \ref{sample_selection}). From bottom to top galaxies show more and more settled and regular photometries.}
    \label{PB_gallery}
\end{figure*}

In this section we show an image gallery of low-\delSD\ galaxies with different central $\rm D_n4000$. Fig. \ref{PB_gallery} shows a montage of galaxies arranged by $\rm D_n4000_{center}$, with younger central stellar populations at the bottom. Galaxies on the left of the orange dividing line are classified as having disturbed morphologies (black triangles in Fig. \ref{PB_subsets}) while galaxies on the right side are from the ``clean'' sample. The younger objects tend to show clumpier and more globally irregular, non-axisymmetric light distributions.
This is consistent with the scenario we propose in Fig. \ref{disturbance_diagram}, and also well-matched by theoretical simulations of merging gas-rich galaxies, which typically show gas flows to the center, resulting in central starbursts \citep{Hopkins2013}.

\bsp	
\label{lastpage}
\end{document}